\documentclass[sigconf]{acmart}
\usepackage{macros}
\usepackage{makecell}
\usepackage{multirow}
\usepackage{tabularx}
\usepackage{graphicx}
\usepackage{soul}
\AtBeginDocument{%
  }

\copyrightyear{2026}
\acmYear{2026}
\setcopyright{cc}
\setcctype{by}
\acmConference[IUI '26]{31st International Conference on Intelligent User Interfaces}{March 23--26, 2026}{Paphos, Cyprus}
\acmBooktitle{31st International Conference on Intelligent User Interfaces (IUI '26), March 23--26, 2026, Paphos, Cyprus}
\acmPrice{}
\acmDOI{10.1145/3742413.3789107}
\acmISBN{979-8-4007-1984-4/2026/03}

\begin{document}

\title[\sysname{}]
{\sysname{}: Supporting Unfamiliar Online Decision-Making with Multi-Agent Conversational Interactions}



\definecolor{color1}{HTML}{C6E6EE} 
\definecolor{color2}{HTML}{E1D7F0} 
\definecolor{color3}{HTML}{FAC5B9} 
\newcommand{\criterion}[1]{\colorbox{color1}{{#1}}}
\newcommand{\option}[1]{\colorbox{color2}{{#1}}} 
\newcommand{\agent}[1]{\colorbox{color3}{{#1}}}

\newcommand{\bryan}[1]{{\color{blue}{#1}}}
\newcommand{\revise}[1]{\textcolor{red}{#1}}

\definecolor{Issue1}{HTML}{ff7e16}
\definecolor{Issue2}{HTML}{4860cc}
\definecolor{Issue3}{HTML}{309f30}
\definecolor{Issue4}{HTML}{d52b2c}
\definecolor{Issue5}{HTML}{9367bc}

\newcommand{\etal}{et al.}

\newcommand{\multiagent}{{\textsf{MultiAgent}}}
\newcommand{\web}{{\textsf{Web}}}

\author{Jeongeon Park}
\orcid{0000-0002-8353-0431}
\email{jep034@ucsd.edu}
\affiliation{%
  \institution{University of California, San Diego}
  \city{La Jolla, CA}
  \country{USA}
}
\authornote{This work was mostly conducted while Jeongeon was at KAIST.}

\author{Bryan Min}
\orcid{0009-0003-0657-4398}
\email{bdmin@ucsd.edu}
\affiliation{%
  \institution{University of California, San Diego}
  \city{La Jolla, CA}
  \country{USA}
}

\author{Kihoon Son}
\orcid{0000-0001-7224-2947}
\email{kihoon.son@kaist.ac.kr}
\affiliation{%
  \institution{School of Computing, KAIST}
  \city{Daejeon}
  \country{Republic of Korea}
}

\author{Jean Y. Song}
\orcid{0000-0003-4379-3971}
\email{jeansong@yonsei.ac.kr}
\affiliation{%
  \institution{Information and Interaction Design, Yonsei University}
  \city{Incheon}
  \country{Republic of Korea}
}

\author{Xiaojuan Ma}
\orcid{0000-0002-9847-7784}
\email{mxj@cse.ust.hk}
\affiliation{%
  \institution{Hong Kong University of Science and Technology}
  \city{Hong Kong}
  \country{China}
}

\author{Juho Kim}
\orcid{0000-0001-6348-4127}
\email{juhokim@kaist.ac.kr}
\affiliation{%
  \institution{School of Computing, KAIST}
  \city{Daejeon}
  \country{Republic of Korea}
}
\email{juho@skillbench.com}
\affiliation{%
  \institution{SkillBench}
  \city{Santa Barbara, CA}
  \country{USA}
}






\newcommand{\sysname}[0]{ChoiceMates}

\begin{teaserfigure}
  \includegraphics[trim=0cm 1cm 0cm 0cm, clip=true, width=0.95\textwidth]{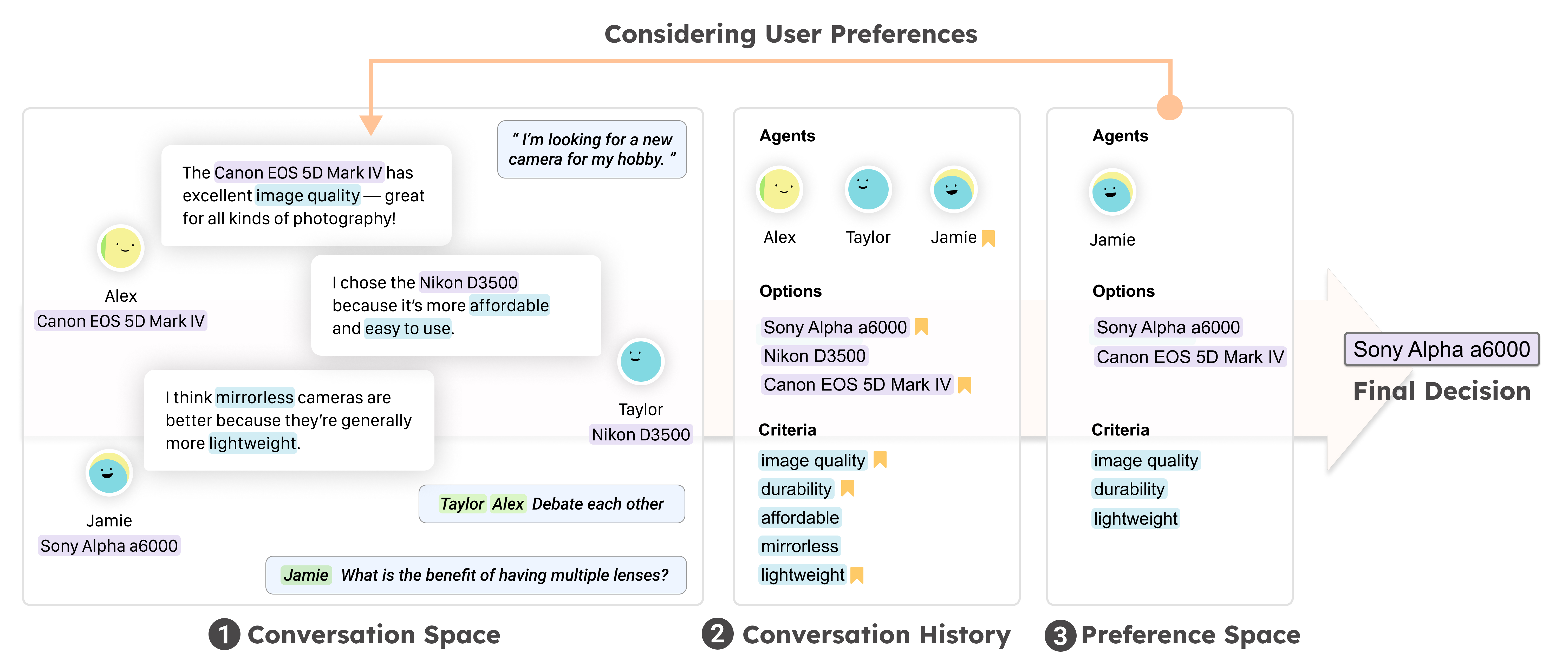}
  \caption{\sysname{} is an interactive multi-agent conversational system designed to support online decision-making in unfamiliar domains. In \sysname{}, the user can converse with any selected set of agents in the conversation space (1) to \emph{gather} diverse information and perspectives in the domain (supported by features that auto-highlight criteria and options in the conversation history) (2) to \emph{manage} and understand the information found. The user can pin important keywords to the preference space (3) to build their own preference in the process and make an informed and confident \emph{final decision}.}
  \Description{
  A flowchart with three sections titled 'Considering User Preferences,' 'Conversation History,' and 'Preference Space' shows a discussion among three individuals, Alex, Taylor, and Jamie, about choosing a new hobby camera. They compare the Canon EOS 5D Mark IV, Nikon D3500, and Sony Alpha a6000 based on image quality, durability, affordability, and weight. The final decision is the Sony Alpha a6000.
  }
  \label{fig:teaser}
\end{teaserfigure}

\begin{abstract}


From purchasing a gift to deciding on a hobby, unfamiliar decisions---decisions without domain knowledge and experience---are frequent and significant. The complexity and uncertainty of such decisions demand unique approaches to information seeking, understanding, and decision-making. Our formative study highlights that in the current workflow, users want to start by discovering broad and relevant domain information evenly and simultaneously, quickly address emerging inquiries, and gain personalized standards to assess information found.
We present \sysname{}, an interactive multi-agent system designed to address these needs by enabling users to engage with a dynamic set of LLM agents each presenting a unique experience in the domain. Unlike existing multi-agent systems that automate tasks with agents, the user orchestrates agents to assist their decision-making process in each turn, through chatting with all agents, with a tagged subset of agents, or calling in new agents into the space.
By comparing \sysname{} with a web search condition and a multi-agent framework (n=12), we show that \sysname{} enables a more confident, satisfactory decision-making with better situation understanding than web search, and higher decision quality than a commercial multi-agent framework. We further illustrate how participants utilized \sysname{} to make unfamiliar decisions, providing insights into designing a more controllable and collaborative multi-agent system.

\end{abstract}



\maketitle

\section{Introduction} \label{introduction}

People make decisions online every day, from purchasing a new product, choosing a trip destination, to taking on a new hobby. In making a decision, people inherently go through a set of steps including recognizing the need for decision-making, collecting information, identifying and ranking key criteria, identifying and evaluating alternatives, and deciding the final choice among them~\cite{lunenburg2010decision}. 
However, making an \emph{unfamiliar} decision---decision-making situations where an individual lacks knowledge and preference in the domain---requires individuals to first grasp the fundamental knowledge of the domain before delving into the decision-making process \cite{Karimi2015-priorknowledgetostyle, bessette2021people}.
Such lack of familiarity makes decision-making more challenging as it requires the user to continually switch between learning about the domain and evaluating options, often under a limited time budget \cite{Bettman1980-phasechoiceprocess, moore1980individual, malhotra1983individual}.


While prior work has demonstrated effective support for online decision-making, they have focused on providing support for managing and organizing criteria and alternatives in the domain rather than for uncovering and understanding them \cite{Chang2020-mesh, Liu2019-unakite, Liu2022-crystalline}. This lack of support during the initial phases of searching for and understanding the key domain information can increase information overload for unfamiliar decision-making situations \cite{iyengar2000choice, scheibehenne2010can}. Subsequently, information overload often leads to a significant investment of time and cognitive resources in grasping domain details and eliciting preferences, potentially causing individuals to make uninformed decisions or even give up entirely \cite{jameson2015human, thai2017too}. Beyond increasing effort, overload can also undermine users' confidence in their judgment and satisfaction with the resulting choice, especially when users must make sense of a new domain \cite{lee2004effect, crescenzi2021adaptation, starke2023examining}.

On the other side, other research aids sensemaking of unfamiliar domains through an interactive and guided exploration of comprehensive knowledge \cite{liu2023selenite}. While such an approach ensures a holistic comprehension of the domain, users still need to put a significant effort into connecting the knowledge to their situations, especially for new decisions.
To minimize user effort, recommender systems aim to simplify decision-making by suggesting personalized options \cite{chen2013human, haubl2000consumer}. Yet, for novices in a domain, these preferences might not be well-elicited \cite{lops2011content, schafer2007collaborative}. This can cause an over-reliance on the system, especially if the novices cannot detect errors due to their limited knowledge \cite{Nourani2020TheRO}. 
Ultimately, novices face the dilemma of researching the unfamiliar domain individually which is demanding, evaluating different options without good confidence and intuition, or ending up making a decision that is ill-informed, all of which motivates the need for a new approach that supports unfamiliar decision-making.

To understand the practice and needs of online decision-making in an unfamiliar domain, we conducted a formative study with 14 participants who had experience making decisions online. From observing and interviewing their decision-making process in unfamiliar domains, we discovered that participants described the need to understand a wide range of opinions and perspectives (including relevant and broad domain information) simultaneously, quickly explore inquiries that are on top of their minds, and develop personalized standards to access the information found.

The identified needs point to a challenge that is inherently about comparing perspectives: novices often need to quickly sample and compare a wide range of viewpoints to learn what matters in a domain, while also asking follow-up questions as new uncertainties arise. This motivated us to explore \textbf{multi-agent conversational interactions}. Multiple agents can surface diverse opinions and perspectives in the domain through distinct viewpoints, allowing users to juxtapose opinions, discover trade-offs, and surface relevant criteria early rather than assembling information from scattered sources sequentially. Conversational interaction can support lightweight exchanges of information needs: users can ask follow-ups, compare options, and focus the search on their context. Together, a conversational set of agents could help users learn key domain knowledge, identify plausible options, and develop personalized standards within a limited time budget.

Many existing multi-agent systems primarily aim to automate the process of solving complex tasks with minimal human intervention by assigning agents into different subtasks \cite{li2023traineragent, yuan2024mora} or roles \cite{wang2024learning, inoue2024drugagent, wang2024learning}. However, there is a limited exploration of multi-agent interfaces with interactions where the user is assisted or empowered by collaborating with a set of agents. For tasks like information seeking and (unfamiliar) decision-making where human preferences and opinions play a significant role, it is important to keep users in the loop, as full automation may lead the users to naively follow agents' guidance and lose agency over their decisions.

We present \textbf{\sysname{}}, a multi-agent conversational system where the user can navigate the unfamiliar decision-making process by conversing with a set of agents and constructing the user's own preference space. In \sysname{}, each agent is characterized by a unique descriptor (i.e., a single-line description of the agent; a professional photographer in Fig. \ref{fig:agent-profile}), a set of valued criteria (quality, durability, and battery in Fig. \ref{fig:agent-profile}), and a single option (Canon EOS 5D Mark IV in Fig. \ref{fig:agent-profile}) to reflect the unique experience, to support intuitive exploration of diverse perspectives. The user takes full control of the agents, and can converse with any individual or group of agents in the \textit{conversation space} (Fig.\ref{fig:teaser}-1) to explore and expand the information in the decision space. Agents, auto-detected criteria, and options from the conversation appear as the \textit{conversation history} (Fig.\ref{fig:teaser}-2) to help users uncover and keep track of the key information found. The user can save agents, criteria, and options to the preference space (Fig.\ref{fig:teaser}-3) with brief notes throughout the conversation to shape the user's preferences and utilize them when making the final decision.

To evaluate whether an interactive multi-agent-based approach can effectively support unfamiliar decision-making, we conducted a three-condition within-subjects user study with 12 participants. We compared \sysname{} with a web search condition and a commercial multi-agent framework condition \footnote{Multiagent Wizard: https://chat.openai.com/g/g-u9C6YeMsL-multiagent-wizard} where a set of necessary agents collaborate to respond to the user, guided by the following research questions:
\begin{description}
    \item[RQ1.] How does \sysname{} help users explore a broad information space in the domain?
    \item[RQ2.] How does \sysname{} support the discovery and management of relevant information to the user’s context?
    \item[RQ3.] How does the user perceive their final decision with \sysname{}?
    \item[RQ4.] How does the user utilize \sysname{} in the decision-making process?
\end{description}

 Our results show that \sysname{} successfully led to a more satisfactory and confident decision, enables comparable exploration of a broad information space, and supports easier management of discovered information and a better understanding of user situation than the web condition. Compared with an existing computer-led multi-agent framework, \sysname{} facilitated broader exploration of information in the domain and participants perceived the final decision as of higher quality. In addition, we report on representative strategies of how participants utilized \sysname{} for unfamiliar decision-making, and further discuss the user-side implications and design considerations of a controllable, collaborative multi-agent system.

This research makes the following contributions:
\begin{itemize}
    \item Findings from a formative study with 14 participants uncovering the practices and needs in unfamiliar decision-making.
    \item Design and implementation of \sysname{}\footnote{The code repository and supplementary materials for \sysname{} is available at https://github.com/jeongeonp/ChoiceMates.}, a multi-agent conversational system designed to support unfamiliar decision-making by providing users control over conversations with a dynamic set of agents with unique experiences and preference construction in the form of criteria, options, and agents. 
    \item Findings from a study with 12 participants that revealed \sysname{} supports discovery of broad information with less burden, effective management of discovered information, and a more satisfactory and confident final decision compared to the \web{} condition, and a wider exploration of information space and a perceived quality final decision with higher confidence compared to the \multiagent{} condition.
\end{itemize}

\section{Related Work}
Our work addresses the challenges in online \textit{unfamiliar} decision-making by leveraging the benefits of multi-agent conversations powered by LLMs. We review prior work on online decision-making support, conversational interfaces for online decision-making, and designs of collaborative multi-agent systems.

\subsection{Online Decision-Making Support}

Due to the enormous amount of information online, decision-makers can face information overload \cite{wan2003happens}, which negatively influences the resulting decision quality \cite{peng2021does, lee2004effect}. For unfamiliar decisions where users do not have sufficient knowledge or expertise in the domain, searching for information becomes particularly challenging \cite{Karimi2015-priorknowledgetostyle, bessette2021people}. 
Beyond increasing effort, overload can undermine users' confidence and satisfaction with the resulting choice, particularly when they must identify what criteria matter and evaluate options in a new domain \cite{lee2004effect, crescenzi2021adaptation, starke2023examining}.

One thread of research focused on identifying selection criteria among a wide range of options to choose from. 
Mesh proposed consumer product decision-making through comparison tables and customizable preference settings \cite{Chang2020-mesh}, and Unakite and Crystalline supported software developers' decision-making with web content snippets and comparison tables \cite{Liu2019-unakite, Liu2022-crystalline}.
While these systems provide strong support for utilizing collected information on options to make decisions, they provide limited assistance in helping users understand information when users lack domain knowledge. Consequently, novices in the domain may not only struggle to identify important criteria but also to critically compare discovered options \cite{iyengar2000choice, scheibehenne2010can}.
On the other hand, Selenite leveraged LLMs to generate potentially relevant criteria based on user search and annotate them in the information inside websites to structure the exploration process \cite{liu2023selenite}. While Selenite ensures a high comprehension of the domain, the users would need to spend additional effort to precisely understand how the criteria match their unique situation to make a decision.

To make the process less overwhelming, recommender systems can guide users unfamiliar with the domain by providing personalized options for users to choose from \cite{chen2013human, haubl2000consumer, Shapira2010RecommenderSH}. 
However, one-shot recommender systems (i.e. recommendations given after just a single example or interaction) are insufficient to support users in learning about the domain for unfamiliar decisions as the interaction is uni-directional \cite{Jannach2021-rg}. This can decelerate users from forming their preferences during the decision-making process \cite{lops2011content, schafer2007collaborative, wang2013research}.
Conversational recommender systems (CRS) build on top of existing recommender systems by offering the ability to make back-and-forth exchanges of messages to continually provide contextualized and relevant recommendations throughout the user's learning and decision-making process \cite{Jannach2021-rg, Wang2013-hd, Warnestal2005-yv}. However, CRS often struggle with maintaining context and providing sufficient explanations that the user need \cite{jannach2022conversational}.

\sysname{} is primarily inspired by the benefits of CRS in providing personalized recommendations and guidance for preference elicitation and aims to expand that to unfamiliar decision-making situations where a holistic understanding of a domain is crucial for making a confident decision. We additionally incorporate the benefits of web-based decision-making support systems to identify and manage key criteria and options in the domain better.

\subsection{Text-Based Conversational Interfaces for Information Seeking}

Text-based conversational interfaces have been widely adopted for various tasks, valued for their ability to present information in a human-like, accessible format and support flexible, adaptive exploration \cite{liao2020conversational, xiao2023inform, xiao2020tell, foo2022papr}.
They have also proven effective in information seeking behaviors online, where Searchbuddies \cite{hecht2012searchbuddies} embedded search engine agents in social media threads to provide easy access to relevant information online, and Gupta et al. \cite{gupta2022trust} increased user trust and satisfaction through conversational online housing recommendations. In addition, Radlinski et al. \cite{Radlinski2017-convsearchframework} proposed a set of properties that composes a natural and efficient conversational information retrieval system.
The recent surge of large language models (LLMs) has revealed the potential for more capable conversational interfaces across diverse domains ranging from programming to searching for UI \cite{wang2023enaling, liffiton2023codehelp, ma2023understanding}. One most prominent example for information seeking is ChatGPT and other LLM applications, where they support real-world tasks such as trip planning or learning a new language \cite{ChatGPT-plugins}. 
However, such interfaces often present information as a linear stream, which can make it difficult to compare options and keep track of evolving criteria, which are critical steps in unfamiliar domains.

On the other hand, while conversational interfaces adaptively provide information through a multi-turn process, conversing requires the user to articulate their input clearly which, if not done, could result in misinterpreted user intent and irrelevant responses \cite{zue2000cuiadvances, Candello2016-designconvui, flohr2021chatortap, jain2018evalchatbots, luger2016badpa}. 
To complement that, existing work has demonstrated techniques to bring multi-modal inputs such as a graphical user interface (GUI) to repair conversational breakdowns \cite{li2020multi, li2021demonstration+}. LLM-based interfaces have also become more proactive, asking clarifying questions and offering candidate options when users’ intents are underspecified.

\sysname{} builds on these LLM-powered conversational interfaces by providing interactive and digestible conversations to support online decision-making. In addition, \sysname{} adds on direct manipulation of objects as an interaction modality on top of natural language, where actions such as saving valuable criteria or options to the preference space are fed into the conversation as a context. 
This moves beyond a linear conversational interaction by letting users externalize and organize information, where these actions become part of the conversational state, enabling more precise and user-steered preference articulation.


\subsection{Multi-Agent Interfaces}
Earlier multi-agent works show that many human tasks benefit from collaboration and information integration from various sources \cite{curcseu2015cognitive, goertzel2017formal} compared to individual cognitive processes, and capitalize on unique identities of each agent and their abilities to delegate tasks among agents \cite{kitamura2002multiple, kitamura2004web, li2023leaders}. 
Synthesizing such investigations with technical advancements, multi-agent systems have recently gained popularity with the emerging capability of Generative AI and LLM. To automate human work, they utilize the interaction between multiple agents and/or with humans to simulate human behaviors \cite{park2022socialsimulacra, park2023generativeagents}, cooperate to solve complex tasks \cite{wang2024learning, li2023traineragent, zolitschka2020novel}, generate visual contents \cite{yuan2024mora, fan2024contextcam, ccelen2024design}, and leverage perspectives from diverse agents \cite{chan2023chateval, cox2023prompting}. Frameworks such as AutoGen \cite{wu2023autogen} or AgentVerse \cite{chen2023agentverse} support this process of building a multi-agent system by helping developers flexibly define and configure agents and conversation patterns.  

Distinguishing from multi-agent systems that aim to automate human work with minimal human intervention (e.g., adjusting input data), \sysname{} aims to design a system that can foster collaboration between multiple agents and humans. 
Prior work explores effective designs of agents or multi-agent conversations to \emph{assist} human judgment and task execution. For example, Benharrak et al. \cite{benharrak2024writer} asked writers to define and identify AI personas that can provide on-demand feedback from different perspectives. ChainBuddy \cite{zhang2024chainbuddy} assisted the planning process by assigning specific agents to each task in the plan, allowing for a focused execution. Moreover, CommunityBots \cite{jiang2023communitybots} evaluated a conversation and topic management mechanism with multiple agents each specializing in a topic for gathering public input across domains and topics with a Wizard of Oz study. While they provide interesting insights into multi-agent interaction design, they offer limited direct evidence on how such designs change users' decision-making processes.

Complementing these, recent HCI work examined how multi-agent conversational interfaces can broaden the set of perspectives users encounter (e.g., to counter filter bubbles) \cite{zhang2024seewidely}, increase intended sustainable behaviors through guidance from multiple NPCs \cite{zhang2024can}, and how multi-agent group dynamics may influence users' judgments \cite{song2025multiagents}.
Building on this growing literature, we investigate the design and implementation of multi-agent interfaces in the context of information seeking and comprehension for unfamiliar decision-making processes, an underexplored domain. Specifically, we design agents that not only provide factual information but also surface diverse perspectives by revealing similar and contrasting experiences and opinions in response to user inquiries, aiming to deepen users' and support more informed decision-making through enriched interactions.

\section{Formative Study}
To design an approach to support unfamiliar decision-making, we first aimed to understand the common practices in unfamiliar decision-making and uncover the needs in the process. For this purpose, we conducted semi-structured interviews with 14 participants with prior experience in unfamiliar decision-making.

\subsection{Participants and Study Procedure}

We recruited 14 participants (Age=18-55, M=28.2, Std=7.8; 8 males and 6 females) who had frequent experience making decisions online (details in Table \ref{tab:formative-demographics}). The recruitment took place through our university’s community channels and snowball sampling to recruit participants in a wide age group. To ensure their experience in making decisions in unfamiliar scenarios, we asked the participants to list their previous experience in both familiar and unfamiliar decisions in the recruitment form.

The interview lasted for 75 minutes for each participant. First, a 20-minute semi-structured interview was conducted to understand participants’ previous experiences with unfamiliar decision-making. We asked questions about the overall process, which information they utilized, and the challenges and needs they faced in the process. Then, a 30-minute think-aloud study was conducted where participants chose an unfamiliar scenario from the given set (e.g., buying a robot vacuum cleaner or choosing a new hobby to do in their free time; see Appendix~\ref{appendix:formative-study-scenarios}) and browsed through the internet until they identified several solid options. \footnote{As we allocated a rather short time for the think-aloud study, we complemented the observation with an interview about their past experiences.} Lastly, participants were interviewed for 15 minutes on their experience with their process and the challenges and needs they faced in this scenario, then on the general challenges and desired support of unfamiliar decision-making. Participants were compensated 20,000 KRW (approximately 15 USD) for the interview. The study was approved by the Institutional Review Board (IRB) at the institution this research was conducted.


After the study, the interviews and think-aloud sessions were transcribed using automated tools and manually reviewed for accuracy by the first author. The two authors then conducted an inductive thematic analysis, independently coding participants' current practices, existing needs, and desired support. The authors collaboratively discussed discrepancies, resolved conflicts through consensus, and iteratively refined the codes to finalize the findings.

\begin{table*}[ht]
\centering
\begin{tabularx}
{0.95\textwidth}{c|c|c|l|m{2.5in}}
\hline
\textbf{ID} & \textbf{Age}   & \textbf{Gender} & \textbf{Previous Unfamiliar Scenario} & \textbf{Selected Scenario} \\ \hline
P1             & 26-35 & F      & \makecell[l]{Choosing a course to take for a summer school, \\ 
Buying an oven for a baking hobby} & Choosing a cafe to cater some snack food on an end-of-semester event \\ \hline
P2             & 36-45 & M      & \makecell[l]{Choosing a trip destination, \\Choosing a t-shirt for summer} & Planning a solo trip destination for three days \\ \hline
P3             & 26-35 & M      & \makecell[l]{Buying a tennis racket, \\ Choosing a U.S. stock option to buy} & Buying a car seat for a friend’s newborn \\ \hline
P4             & 26-35 & F      & \makecell[l]{Choosing a working holiday location, \\ Buying a laser hair removal} & Buying a car seat for a cousin’s newborn \\ \hline
P5             & 18-25 & M      & Buying a lunch box & Buying a robot vacuum cleaner to replace the normal vacuum cleaner \\ \hline
P6             & 26-35 & F      & \makecell[l]{Buying a new tennis racket, \\Choosing meal to cook} & Renting a house short-term for an internship \\ \hline
P7             & 18-25 & M      & \makecell[l]{Choosing a transportation card plan \\in a new country} & Choosing a new hobby to do in free-time \\ \hline
P8             & 18-25 & F      & \makecell[l]{Planning a trip to Osaka, \\ Choosing a wrist/ankle brace+lumbar support} & Buying a robot vacuum cleaner to replace the normal vacuum cleaner \\ \hline
P9             & 18-25 & M      & \makecell[l]{Choosing a hair salon, \\ Buying a gift for a friend} & Buying a car seat for a friend’s newborn \\ \hline
P10            & 18-25 & M      & \makecell[l]{Whether to get a CT+MRI during ER visit, \\ Buying clothes at a flee market} & Buying a skateboard for transportation purposes instead of walking to school \\ \hline
P11            & 26-35 & F      & \makecell[l]{Buying a coffee pod machine, \\ Searching for a job} & Buying a skateboard for transportation purposes instead of walking to school \\ \hline
P12            & 26-35 & M      & \makecell[l]{Buying a used phone, \\ Planning a trip to Jeju} & Buying an interior light at home \\ \hline
P13            & 18-25 & M      & \makecell[l]{Choosing an affordable phone plan, \\ Choosing field of work} & Buying a skateboard for transportation purposes instead of walking to school \\ \hline
P14            & 46-55 & F      & \makecell[l]{Choosing exercise, \\ Buying a blue light blocking film} & Buying an interior light at home \\ \hline
\end{tabularx}
\caption{Participant demographic for the formative study.}
\Description{A table showing participant demographic for the formative study, with the following information in each of the columns: ID, age, gender, previous unfamiliar scenario, and the selected scenario. There is a total of 14 participants.}
\label{tab:formative-demographics}
\end{table*}

\subsection{Findings}
Here, we describe the different approaches people follow for unfamiliar decision-making and describe the three main needs we identified from the process.

\subsubsection{Practice}
The participants utilized several sources in an unstructured manner (i.e., in various orders and frequencies) to collect information and make unfamiliar decisions. 
Such sources include summary posts of the domain (n=8), posts with individual opinions and experience (n=10), and websites with a list of options in the domain (n=11). During the interview, many (n=12) brought up asking an expert as their previous experience to get helpful and needed information for making decisions in unfamiliar domain. 

With \emph{summarized content} such as YouTube summarization videos or rating sites (e.g., https://www.rtings.com/), participants were able to objectify multiple criteria and understand an overview of the domain, but could not receive information more relevant to their context. Participants found \emph{posts with individual opinions}---online communities, product reviews, or blog review posts---helpful for gathering diverse opinions on the options but also mentioned that it is time-consuming to find credible sources and the possible bias hinders them from utilizing individual opinions solely.
\emph{Browsing through a grid of options} from e-commerce websites (e.g., https://www.amazon.com/) was another approach participants took, but they were unable to understand and compare different options without knowledge of the domain, thus switching to the other two sources of information. 

On the other hand, 12 out of 14 participants mentioned \emph{asking an expert}---including store managers, friends, and family who have expertise in the decision domain---an alternative approach they took for past unfamiliar decisions as a way to receive information without being overwhelmed. They mentioned that experts' ability to (1) explain ground-based (i.g., basic but essential) information about the domain and provide information that’s difficult to find, and (2) ask clarification questions to make better suggestions in one’s specific context example makes them reach out to experts for unfamiliar decisions.

\subsubsection{Needs}
We identified three main needs participants exhibited during the unfamiliar decision-making process.

\paragraph{C1: Discovering broad and relevant domain information evenly and simultaneously}
The participants mentioned two major types of information that assist their unfamiliar decision-making process, namely broad (i.e. information that represents comprehensive information of the domain) and contextually relevant (i.e. information that is relevant to one's situation or preferences) domain information. Participants expressed concerns about `missing out on the core domain information' (P14) but were also frustrated when they were `not able to find information relevant to their case' (P5). As the domain was unfamiliar, participants often ended up exploring one type of information and struggled to start exploring the other type of information. Participants tried to mitigate this by expanding their discovery to more fresh experiences or opinions in the domain, to ensure that they understood the landscape of the domain. However, due to their lack of ability to discover information in an unfamiliar domain, many could not diversify their search to underexplored types of information.

\paragraph{C2: Quickly addressing inquiries that are on the top of their minds}
As participants had insufficient domain knowledge, they had many inquiries and curiosities during the process. While such inquiries were often quick and short (e.g., difficulty of installing a discovered car seat (P9)), participants deliberately looked for answers instead of skipping as it was a crucial step to establishing a deeper domain understanding. However, their inquiries were not sufficiently addressed due to a lack of time or ability to find what they wanted. 
For example, P12 (interior light) while watching a YouTube video on famous lamp brands, had to go on Google multiple times to search for new information, such as the meaning of Handwerker or mushroom lamp styles, which they later described as tedious. P12 additionally expressed the need to chat with an expert to understand possible materials and their pros and cons.
Similarly, P2 (solo trip destination) mentioned that they would want the person who wrote a particular blog post to answer a few extra questions, including suggestions for other trip destinations and why. 

\paragraph{C3: Gaining personalized standards to assess the information found}
A core concern raised by the participants was that they did not have enough knowledge or confidence in the domain to assess the information discovered. Even after participants read through a few webpages or had notes written down, they still thought they did not have sufficient understanding of their situation to establish personalized standards and evaluate different criteria and options. Thus, participants had to gather a personalized set of key aspects and their importance to assess newly discovered information or make a choice. For instance, P1 (cafe for catering) faced difficulties in ranking the discovered aspects (e.g., location. type of desserts, catering experience) and making the decision until the end when they realized that they were prioritizing cafes with catering experiences.

\section{Design Goals} 
With the identified needs in the unfamiliar decision-making process, we propose the following design goals for a system that can support unfamiliar decision-making.
\begin{itemize}[leftmargin=3em]
    \item[DG1.] \textbf{Provide diverse opinions and perspectives in the domain through experiences.} In unfamiliar decisions, participants wanted to discover a wide range of information, namely broad, relevant, and fresh information with equal focus. Given difficulty identifying what’s ``core'' and answering emerging questions, participants turned to multiple experiences to fill gaps and validate claims. Past research has shown that social information search containing people's experiences could help discover increased firsthand information and diverse perspectives with value judgments \cite{jeon2013value, agrawal2015whither}. In addition, prior work has shown that novices preferred option-based suggestions of preferences over attribute (i.e., criteria)-based suggestions \cite{knijnenburg2009adaptivePE}. Inspired by such works, we can design a system where multiple experiences of using different options can offer complementary perspectives without overwhelming users with unstructured viewpoints.

    \item[DG2.] \textbf{Provide options to explore broad and contextually relevant domain information.} Participants also mentioned the need for simultaneous exploration of broad and relevant domain information to expand the depth and width of their domain knowledge. Support for users to seamlessly switch between the two types of information when needed in the process may help them fill in underexplored parts of their information discovery in the domain.
    
    \item[DG3.] \textbf{Allow for direct exchanges of information needs through conversational interactions.} Conversation is an effective medium for an enhanced searching experience \cite{liao2020conversational} and in expressing information goals \cite{schneider2023investigating}. To aid users in addressing their unique information needs, an effective approach would be to utilize conversational interaction where users can ask follow-up questions on top of the acquired information. This can support further exploration of how different domain information connects to each other and to the user's preferences.
    
    \item[DG4.] \textbf{Facilitate users in building preferences relative to the domain.} The participants mentioned the difficulty of gaining personalized standards (e.g., a travel destination less than 3 hours via flight) that would help them assess the information found. Rather than assuming users can form standards from scratch, the system can use exploration history as a \emph{starting point} to surface preference candidates, and have the users confirm or revise them. To reduce the risk of filter bubbles, the system should also surface contrasting opinions alongside emerging preferences.
    Identified aspects could then function as evidence for the user to evaluate later discovered information and gradually clarify their preferences over the decision-making period.

\end{itemize}

\section{\sysname{}: A Multi-Agent Conversational System}

From our findings and the design goals, we designed and developed \textbf{\sysname{}}, an interactive multi-agent conversational system that assists users in unfamiliar decision-making---situations where users must make a decision in a domain where they lack prior domain knowledge and well-formed preferences, often under a limited time budget. Note that the scope of unfamiliar decision-making \sysname{} is designed to support does not include primarily high-stakes or overly complex decision contexts (e.g., those requiring extensive verification, long-term deliberation, or multi-stage planning), which we discuss further in Section \ref{discussion-3}.

\sysname{} allows the user to (1) converse with a dynamic set of agents to explore broad and relevant information and (2) utilize extracted conversation history to fill up the preferences space to make a decision.
Here, we describe an envisioned user scenario, explain different system components, and describe the system implementation and prompt engineering.

\subsection{Envisioned Scenario}

Sally has recently found an interest in photography and wants to find the right camera for her. However, she is new to the camera domain and has no idea which camera she would be interested in. As she has no experts to consult and gets overwhelmed by doing a few online searches, Sally enters \sysname{} and types into the message input box, ``I’m new to photography, and I want a camera. Not sure which one would be best for me.'' She then encounters three distinctive agents (Alex, Jamie, Taylor) appearing on \sysname{}, each with a unique profile containing an \option{option} (i.e., a camera model) they had chosen and the \criterion{criteria} (e.g., portability, brand) that led to their choice. In each agent's response, she sees criteria and options auto-detected and listed in the conversation history.

\begin{itemize}
    \item Alex: ``As a professional photographer, I value \criterion{image quality} and \criterion{durability} in a camera. That's why I chose the \option{Canon EOS} \option{5D Mark IV}. It’s a full-frame DSLR that delivers excellent image quality and is built to last.''
    \item Jamie: ``I’m a hobbyist photographer and I prefer a camera that’s \criterion{lightweight} and \criterion{easy to use}. I've been using the \option{Sony Alpha} \option{a6000} and it’s been great for me. It’s compact, takes great photos, and is very user-friendly.''
    \item Taylor: ``I’m a travel blogger and for me, \criterion{portability} and \\ \criterion{battery life} are key. I use the \option{Fujifilm X-T3}. It's compact, has a long battery life, and takes amazing photos. What are your needs and preferences when it comes to photography?''
\end{itemize}

Sally learns about the wide range of values and criteria agents with different professions and lifestyles have for cameras through the agents and their profiles (\textbf{DG1}). Sally considers Jamie to be the most relatable as she is getting a camera for a hobby. Sally tags Jamie and asks why they value lightweight and easy-to-use cameras. Jamie responds that they ``don’t need all the bells and whistles of a professional camera'' and they ``want to focus on capturing the moment''. Sally learns that for her personal use, she may want to follow Jamie’s reasoning as Sally is also looking for a simple camera. She then asks follow-up questions to Jamie on ``What is considered a lightweight camera?'' and ``What is the main difference between a professional camera and an easy-to-use camera?'' to better understand what each criterion means (\textbf{DG3}). Sally's conversations with Jamie allow her to discover what is relevant to her context.

After gaining some information, she decides to get an easy-to-use camera and saves the criterion \criterion{easy-to-use} and the agent Jamie to her \textit{preference space}. Then, she turns on the \textit{preference toggle} as she now wants more relevant information to her identified preferences (\textbf{DG2}). As the agents have access to Sally's preference space, when she asks a question that reflects her preference: ``Are there other cameras that I would like?'', three more agents with the criterion \criterion{easy to use} included in their valued criteria appear: Riley, Morgan, and Casey.
Continuing the conversations with different agents, Sally is suddenly lost on which information she needs to discover more. She takes a look at the conversation history, sees the criterion \criterion{durable} mentioned the most, and realizes that she may consider durability a key aspect of her preference. She then adds \criterion{durable} to the preference space and continues exploring the options with relation to their durability (\textbf{DG4}).

As Sally continues to converse and learn more, she becomes more and more confident about the set of criteria she values for her activities and adds those to her preference space (\textbf{DG4}). Now that she has a clearer picture of the domain, she also wants to compare the three cameras she is considering in-depth.
By asking the agents to `debate each other', Sally understands the crucial differences between the options and concludes that \option{Nikon Coolpix B500} best fits her preferences and adds it to her preference space. She checks her \textit{preference space} and the saved criteria and options one more time and confidently makes the decision.


\subsection{System Components}

\begin{figure*}[htb!]
    \centering
    \includegraphics[width=\textwidth]{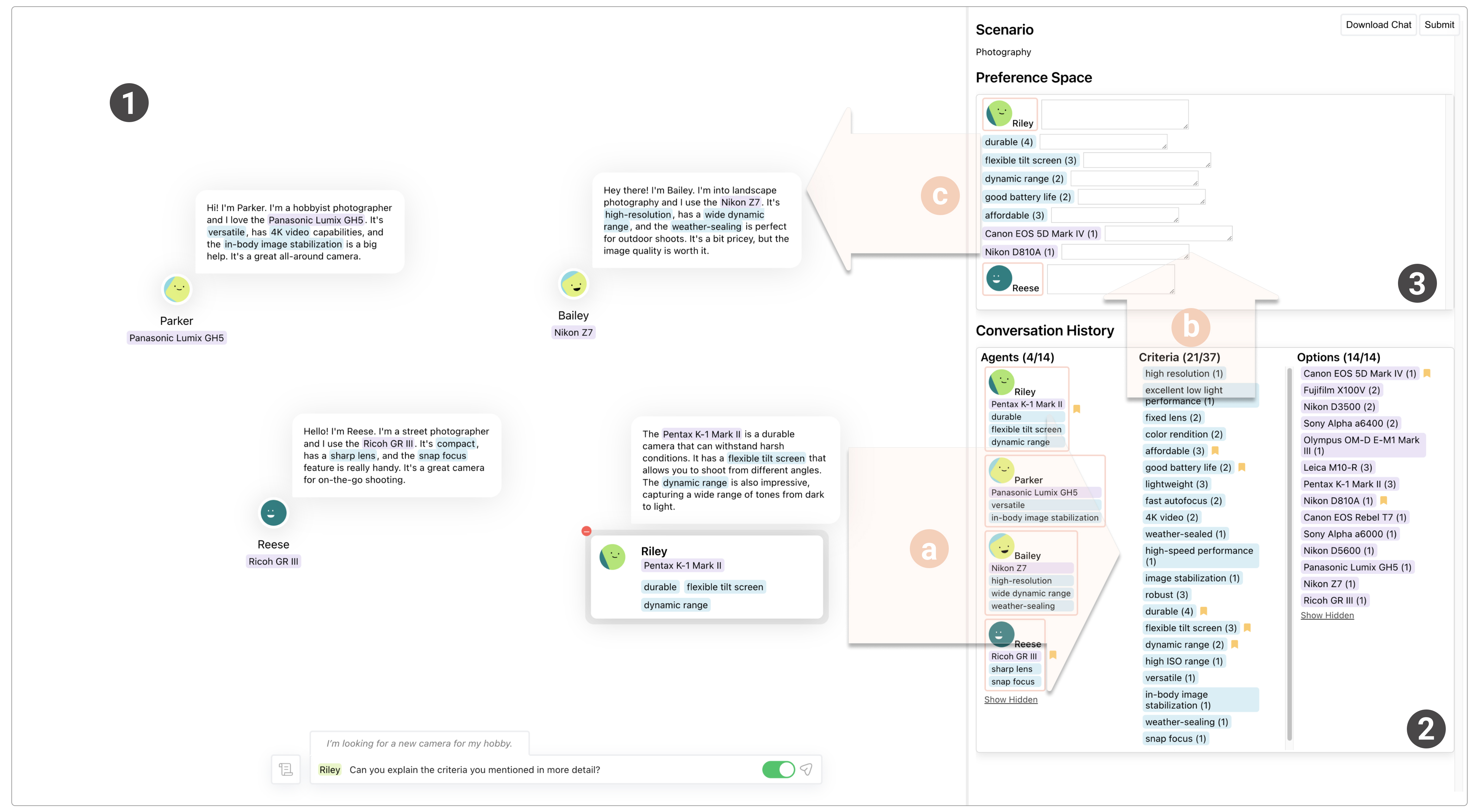}
    \caption{The \sysname{} interface: (1) 
    Agents populate the conversation space to converse with the user, identifying key criteria and options in their utterances (1). The \emph{Conversation History} automatically lists those criteria and options mentioned by agents (2). \emph{Preference Space} contains the options and criteria selected by the user from the Conversation History. (a) indicates the unfamiliar information stacked in the Conversation History. (b) shows the saving process of the user preference. Lastly, (c) illustrates the flow of applying user preference into the Conversation Space with the agents in \sysname{}.}
    \Description{A screenshot of the overall system interface of ChoiceMates. On the left is the conversation space with agents, which shows multiple agents explaining the specific options and important criteria of the camera. On the right, there are two main sections, Conversation History and Preference Space. In the Conversation History, the conversation data with the agents are automatically stacked by Agents, Criteria, and Options. In the Preference Space, there are two agents, five criteria, and two options that the user saved from the Conversation History.}
    \label{fig:interface}
\end{figure*}

\sysname{} (Fig. \ref{fig:interface}) is an interface that allows the user to interact with a set of agents to make an unfamiliar decision. Agents are the main unit of conversation in \sysname{} and are the basis of gaining information about the domain. 

\sysname{} consists of a \emph{conversation space} where the user can converse with a dynamic set of agents, a \emph{conversation history} that automatically lists agents, criteria, options discovered during the conversation, and a \emph{preference space} where the user can gather and save relevant information throughout the process.
In this section, we first describe agents, the conversation space with the agents and additional support in the conversation space to manage the information, then the conversation history and the preference space.

\subsubsection{Agent: Basic unit of information}

\begin{figure}[htb!]
    \centering
    \includegraphics[trim=0cm 0cm 0cm 0cm, clip=true, width=0.45\textwidth]{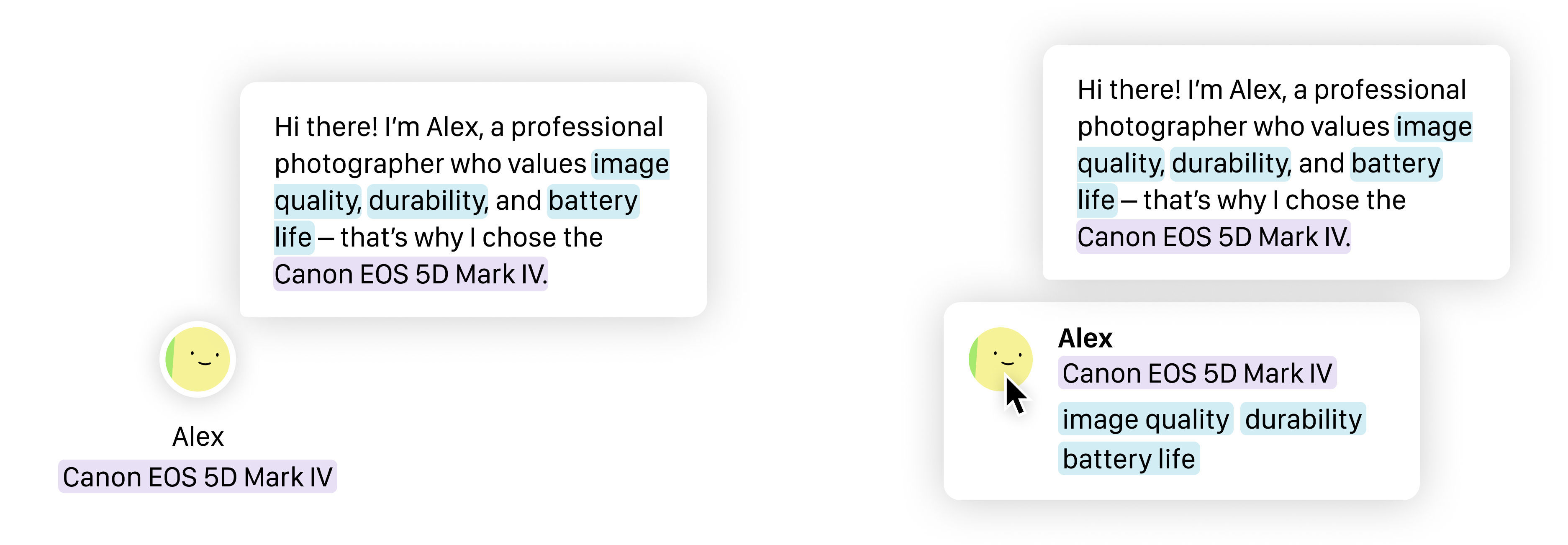}
    \caption{(1) An agent utterance is represented as a chat bubble above the agent icon. (2) Hovering over the icon reveals the agent's profile containing their valued criteria and their chosen option.}
    \Description{The figure shows an example of an agent in the system. On the left is Alex, who is represented by an icon, name, option, and utterance with the criteria highlighted in blue and its option highlighted in purple. The left side shows a cursor hovering over Alex's icon, revealing its profile containing its valued criteria and options.}
    \label{fig:agent-profile}
\end{figure}

In \sysname{}, each agent (Fig. \ref{fig:agent-profile}) is characterized by its descriptor (i.e., a single-line description of the agent; a beginner photographer who likes natural scenes), a set of criteria (i.e., factors to consider in the domain) they value, and a single option (i.e., an available choice in the domain) they chose with the criteria. 
The agents in \sysname{} are designed to each reflect an individual's choice and their underlying values and experience in real life, inspired by previous work showing that novices preferred case-based preference elicitation in a recommender system \cite{knijnenburg2009adaptivePE}. Multiple agents as a group provide users with a broad range of experiences in the space (\textbf{DG1}).
To provide a better salience of criteria and options in the domain, all criteria mentioned by agents are highlighted in \criterion{blue} and all options are highlighted in \option{purple}.

When the user asks an initial question in the domain, \sysname{} prompts the LLM to generate a set of diverse agents, between 3 and 6 to provide a good number of different agents without choice overload \cite{guo2022can}.
Newly spawned agents always begin with introducing themselves, sharing their valued criteria and a chosen option (Messages in Fig. \ref{fig:agent-profile}). To ensure that the agents communicate the correct information, we use a simplified RAG framework (details in \ref{sec:techeval}), by scraping factual information of each option. The scraping happens \textit{right after} each agent is generated to let the LLM produce diverse agents without being restricted by the web-searched information.
The scraped information is provided to the agents and remains in the conversation stream with a prompt asking agents to utilize the information as context.
The user can access the original source of the provided information via the hyperlink added to the option and validate the information throughout the session.

\subsubsection{Conversation Space}

\begin{figure*}[htb!]
    \centering
    \includegraphics[trim=0cm 0cm 0cm 0cm, clip=true, width=\textwidth]{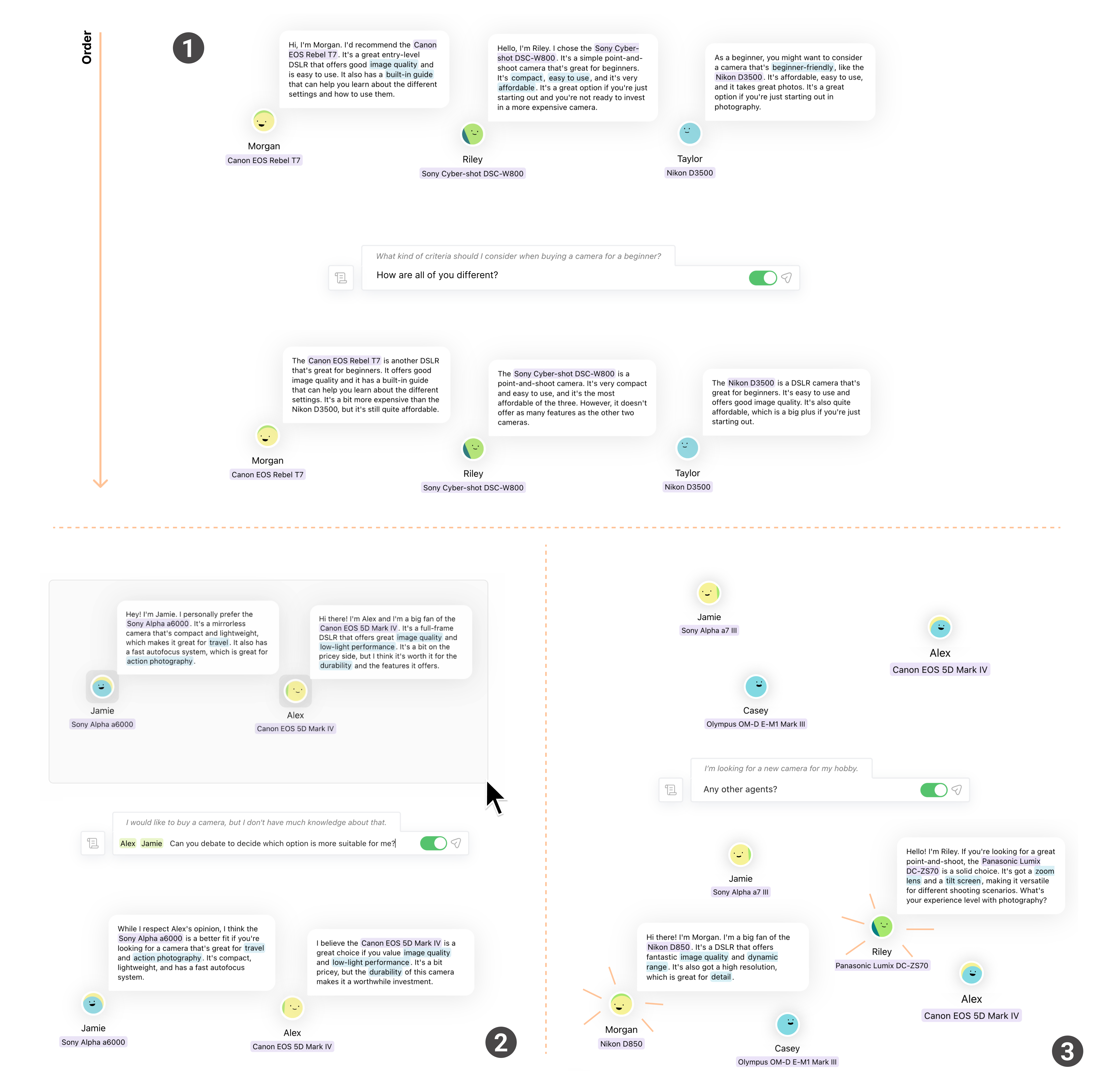}
    \caption{Three types of conversation interactions possible in the conversation space. (1) The user can chat with any agents currently in the space, (2) select or tag one or more agents to chat, and (3) call new agents into the conversation space with their input. Like the conversation in (2), the agents will build on top of each other's responses when applicable.}
    \Description{The figure shows the three main interactions in ChoiceMates. The first interaction illustrates that the user asked the three agents ``How are all of you different?''. Then, the agents explain the differences of the options they suggested. On the bottom left, there is a second interaction when the user asks two agents to debate to decide which option is more suitable for the user. There is a mouse cursor dragging the two agents. On the bottom right, user asked ``Any other agents?'' to see different options. Then, the new two agents pops up with the new options and criteria.}
    \label{fig:conversation-space}
\end{figure*}

To support easy exploration of broad and personal information in an unfamiliar domain, \sysname{} provides a conversational space where all agents reside, where users can converse with a select set of agents (Fig. \ref{fig:conversation-space}-2).

On \sysname{}, agents are designed to chat back and forth with the user (\textbf{DG3}). This way, users can ask follow-up questions to agents they relate to and gain a deeper understanding of domain options.
The user initiates the conversation by sharing their decision-making scenario, and three to six agents with varying profiles appear on the screen. Agents then share their experiences and ask questions about the user's preferences. The user can respond to agents through the message input box by either replying to a question or asking one themselves.

\sysname{} offers both user-driven and system-driven options for selecting agents to respond.
Upon receiving the user's message, \sysname{} detects the user's intent and makes existing agents, new agents, or a combination of both to respond, depending on the agents' relevancy to the user's message and context (More described in Fig. \ref{fig:technical-pipeline})
If the user wants to manually assign particular agents to respond, they can also select agents in the space (Fig. \ref{fig:conversation-space}-2). 
Additionally, agents can converse with one another by responding to other agents in the conversation. They can agree, disagree, or ask further questions to other agents (Fig. \ref{fig:conversation-space}-3). This can reveal comparisons between options and new criteria that the user may not have thought of when they are the only ones conversing with them.
Just as the user can manually select agents to respond, the user can also manually trigger inter-agent conversations by selecting agents and asking them to ``debate each other''.

To reduce clutter from more agents and messages, we chose to only display the latest messages sent from the previous turn. Instead, users could click the button left of the message input box in Fig. \ref{fig:interface} to view the full conversation history as a linear thread.


\subsubsection{Conversation History}
To support a more effective exploration and management of information throughout the conversation, we provide a \textit{conversation history} (Fig. \ref{fig:interface}-2) where agents, criteria, and options are automatically appended as the conversation proceeds. The conversation history also serves as a midpoint to help users transform discovered information into user preferences.
For criteria and options (hereafter keywords), the total count of each keyword is added next to each keyword.

The user can perform three actions in the conversation history. 
To identify the connections between agents and keywords to make better sense of the information in the domain, the user can hover over the keywords to see matching agents and options (for criteria-hovering)/criteria (for option-hovering). Depending on how cluttered the conversation space is, users can hide/unhide agents and keywords. Finally, the user can pin agents and the keywords to add them up to the preference space.

\subsubsection{Preference Space}

\begin{figure*}[htb!]
    \centering
    \includegraphics[trim=0cm 0cm 0cm 0cm, clip=true, width=\textwidth]{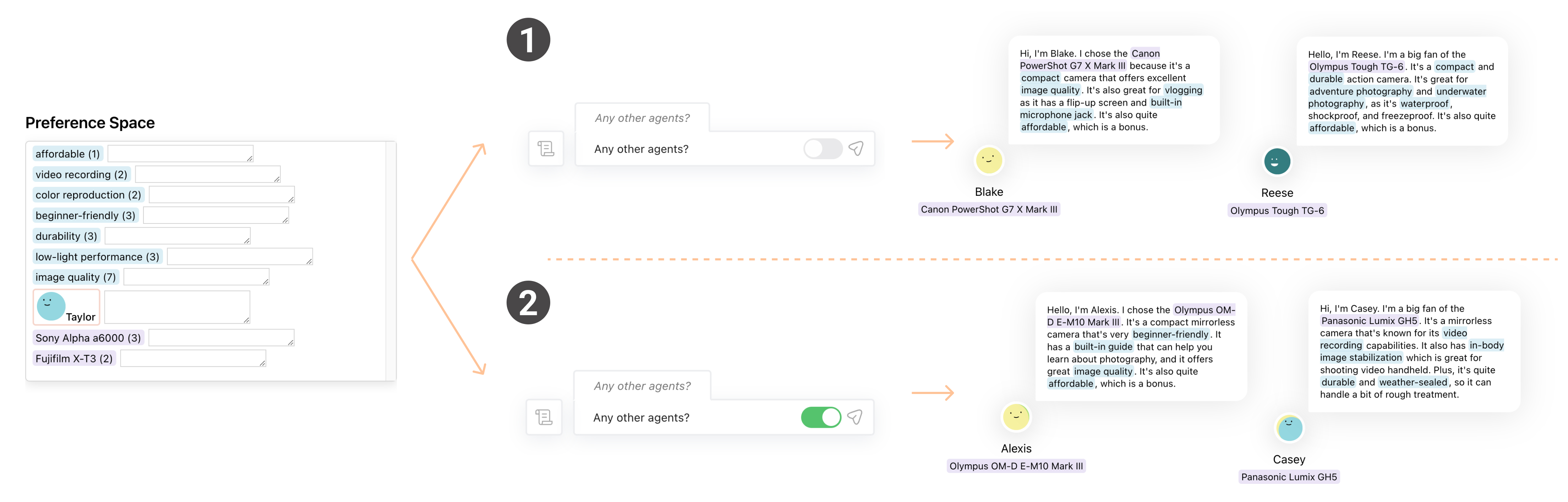}
    \caption{Consequences of preference toggle. With the same user message ``Any other agents?'', when the toggle button is off (1), information in the preference space is hidden from the agents, thus making the responses independent from the user's preferences. When the toggle button is on (2) the responses are tailored to the user's preferences.}
    \Description{This figure explains the difference in whether the toggle button is on or off. By considering the user's preference for the left side, when the toggle button is on, the agents provide the options aligned with the user's preferred criteria. However, when the toggle button is off, the agents provide different options with different criteria.}
    \label{fig:preferences}
\end{figure*}

The \textit{Preference space} (Fig. \ref{fig:interface}-3) provides a dedicated space for the user to store their preferences throughout the session, in the form of agents, criteria, and options. This space supports the user to not only build up their preferences in the domain (\textbf{DG4}), but also to guide the conversations with the agents. 

We use the preference toggle button right to the input message box to give users the option to decide their exploration path between broad and personalized information (\textbf{DG2}, Fig. \ref{fig:preferences}). Turning off the toggle button does not reveal any current preferences to the agent, which makes the agents-to-respond and the responses independent from the user's preference space (Fig. \ref{fig:preferences}-1). On the other side, turning the button on will make the agents-to-respond and the responses more relevant to the user's preference space.


\subsection{Implementation and Prompt Engineering}

\sysname{} was built as a web application using the ReactJS framework, and OpenAI's GPT-4-0613 API was utilized to generate the conversations. To scrape information about the options, we used the Newspaper3k library\footnote{https://newspaper.readthedocs.io/en/latest/} on top of a Flask-based server. The messages and interaction logs were stored in a Firebase realtime database.

\begin{figure*}[h!]
    \centering
    \includegraphics[trim=0cm 0cm 0cm 0cm, clip=true, width=\textwidth]{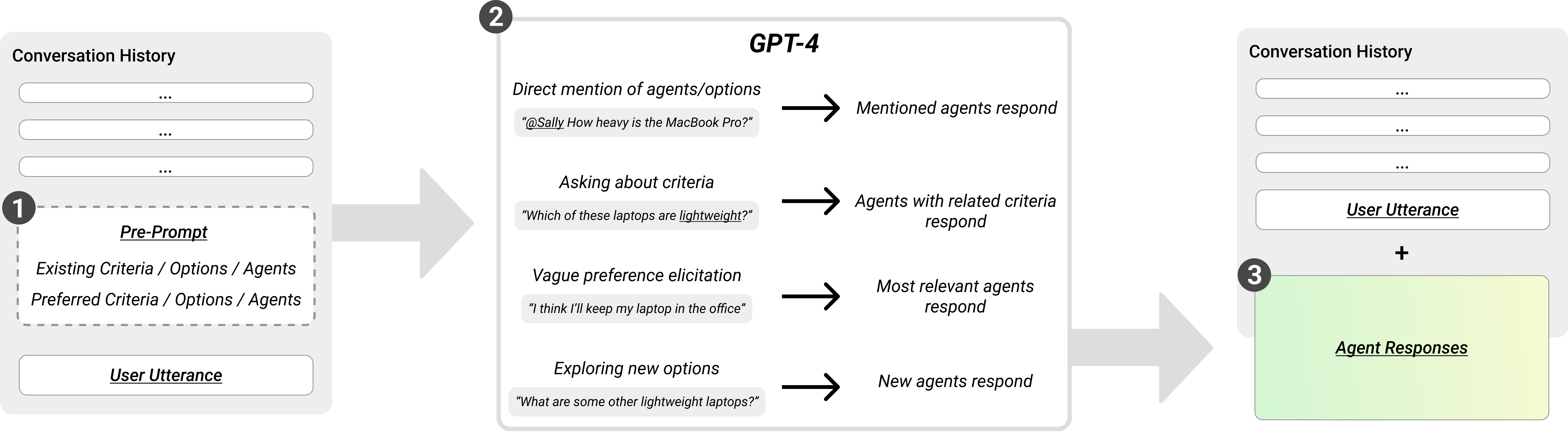}
    \caption{Overview of the technical pipeline. When the user sends their message to \sysname{}, the conversation context is sent to the LLM (1) and is instructed to infer the user's intent (2) and respond through the most relevant agents (3).}
    \Description{The figure illustrates the technical pipeline of ChoiceMates. (1) the conversation context is sent to the LLM with the user utterance as the prompt. The LLM is instructed to identify the type of user utterance: (1) direct mention of agents/options, (2) asking about criteria, (3) vague preference elicitation, and (4) exploring new options. Depending on the user intent, the LLM determines which agents to respond closest to the criteria and options mentioned, or invites new agents to speak up. The agent response is added to the conversation history, but the pre-prompt is not.}
    \label{fig:technical-pipeline}
\end{figure*}

We describe the technical pipeline (Fig. \ref{fig:technical-pipeline}) on managing conversation of multiple agents, and discuss prompting techniques implemented in \sysname{} to facilitate effective multi-agent conversations \footnote{Note that techniques used reflect the time of the tool development and user study, which was between late 2023 and early 2024.}.

The code repository and supplementary materials for \sysname{} is available at https://github.com/jeongeonp/ChoiceMates.

\paragraph{Information Scraping} \label{sec:techeval}
\sysname{} utilizes a simplified RAG framework \cite{lewis2020retrieval} by first searching for the top 3 sites through Google search with the option name, scraping full websites with the Newspaper3k library, then using its output in the prompt context. Our choice of using such a pipeline was due to the technical constraints of the LLM where a larger context could lead to increased hallucination \cite{liu2024lost} and slow real-time response, by keeping a smaller, summarized context.

\paragraph{Prompting for Single-Stream Multi-Agent Conversation}
\sysname{} instructs the LLM to embody multiple personas (i.e. profiles) and manage them simultaneously. The LLM is prompted to embody an identity that communicates through multiple different personas where they are allowed to respond to the user's utterance or any agent's utterance. This instructs the LLM to virtually provide multiple, contextually aligned responses in a single turn of prompting.

\paragraph{Context Retention}
While a linear conversation model makes for a simple data structure to manage, the trade-off of increased reliance on the LLM's ability to retrieve information can cause a loss of context from excess information \cite{liu2023lost}. To address this potential limitation, we prepend a \emph{pre-prompt}---a preliminary, ephemeral prompt message to every prompt of the user's utterance (Fig.~\ref{fig:technical-pipeline}-1).
The pre-prompt contains the lists of criteria, options, and agents currently in the conversation space, and the lists of user-focused criteria, options, and agents. This provides salience to the relevant keywords in the user's decision-making domain. Any action that updates any of the lists also updates the pre-prompt, and once the user sends a message the latest pre-prompt state is prepended.
This technique allows \sysname{} to keep a summary of the conversation state while the conversation history only contains agent messages. While state-of-the-art NLP summary features such as LangChain's Contextual Compression \cite{langchain2023compression} compress large bodies of text into summaries for improved information retrieval, our method utilizes GPT's trait of retrieving information at the beginning and end of the conversation better than in the middle \cite{liu2023lost}.

\paragraph{Representing Multi-Agent Responses}
\sysname{} implements a constrained prompting technique, where the LLM is prompted in such a way that generated text contains interleaved characters to denote tags to guide structure in the textual representation. Inspired by Graphologue's \cite{Jiang2023graphologue} technique, \sysname{} annotates agent names, criteria, and options in the generated text and is parsed by the system interface.



\section{User Study}
\subsection{Setup}

We conducted a user study to understand how effectively \sysname{} supports the unfamiliar decision-making process. The user study was designed as a within-subjects study comparing \sysname{} with two other conditions --- conventional web search and a multi-agent framework.
Our study was designed to address the research questions introduced earlier in Section \ref{introduction} (RQ1-RQ4).

\subsubsection{Participants}
We recruited 12 participants (Age: 9 between 18-25 and 3 between 26-35; Gender: 6M and 6F) who had multiple experiences making decisions online through an online community of the author's university. Most participants (6 `1-2 times a week', 5 `3-4 times a week or almost everyday') indicated that they make online decisions regularly. The participants also varied in their experience utilizing LLMs for recommendations or decision-making (M=3.03, SD=1.56; 1: Never used, 5: Always use). To ensure that the participants were not familiar with the decision domains for the study, we asked them to indicate their familiarity with the 10 selected domain candidates in the recruitment form and the researchers chose the three most unfamiliar ones. We also ensured that the participants were sufficiently fluent in English, as \sysname{} was designed in English. The participants were compensated 40,000 KRW (approximately 30 USD) for two hours of participation. The study was conducted through Zoom~\footnote{https://zoom.us/}, where the participants were asked to prepare a computer or an equivalent device with audio, video, and screen share settings. The study was approved by the Institutional Review Board (IRB) at the institution this research was conducted.

\subsubsection{Conditions}
We compared \sysname{} with two other conditions, a conventional web search baseline (hereinafter \web{}) and an existing multi-agent framework (hereinafter \multiagent{}).  
Each participant used all three interfaces to make three unique unfamiliar decisions. The ordering of the conditions was counterbalanced.

In the \web{} condition, the participants could freely explore the web, including video-based information (e.g., YouTube) and commercial single-agent LLM chat tools (e.g., ChatGPT).
This was designed to resemble a conventional way of making unfamiliar decisions online. We chose this open-ended baseline to reflect how people typically triangulate across sources, including search and single-agent LLM chat tools, rather than relying on a single source.

In the \multiagent{} condition, we used a custom GPT named ``Multiagent Wizard''\footnote{Multiagent Wizard: https://chat.openai.com/g/g-u9C6YeMsL-multiagent-wizard} as a representative multi-agent framework, where the `wizard' automatically creates new agents for specific tasks, and allows them to collaborate to complete tasks.
We included \multiagent{} as one of the conditions to observe the strengths and limitations of existing multi-agent frameworks in supporting decision-making\footnote{While there were more customizable multi-agent frameworks such as AutoGen or Crew, we chose custom GPT over existing frameworks as it was a widely accessible chat-based multi-agent framework to laypeople by the time the study was conducted (March 2024).}.
Among existing multi-agent frameworks available for use, we selected the Multiagent Wizard for its low learning curve being a ChatGPT interface, and for its characteristic of automatically creating the agents to support the process---which is a common design for many existing multi-agent frameworks.
In \multiagent{}, the participants were restricted from accessing other websites.
We did not include a separate single-agent-only condition because it would add a fourth within-subject decision task and increase participant fatigue.

For both conditions, we additionally provided a preference space via Google Docs for the participants to save their preferred criteria, options, and thoughts along the process. We asked them to keep the space open on the side during the sessions to match the preference space in \sysname{}.

\subsubsection{Procedure}

We selected three decision domains for the study: purchasing climbing shoes, a fabric shaver, or a robot vacuum machine, where each participant experienced the same three domains. The domains were equally assigned to the conditions. We employed counterbalancing to ensure that an equal number of participants were assigned to each condition-domain pair. The study lasted for 2 hours, and consisted of the following parts:

\paragraph{Introduction \textit{(10 minutes)}.} The participants were first provided a brief introduction to the study and the scenario assigned. They were then asked to fill out a pre-survey containing questions on a 7-point Likert scale on their confidence in the decision \cite{collier2012conflict}. Afterward, they were introduced to the decision domains and their order.

\paragraph{Decision-making tasks \textit{(30 minutes per condition)}.} Each participant performed three unfamiliar decision-making tasks in the assigned order. For each decision-making task, the participants were first provided with a tutorial on the interface. Then, they were given a maximum of 20 minutes to ``\textit{use the interface, fill in the preference space, and narrow down to a single strong option in the decision space}'' \footnote{We used the term `strong option' to indicate that the decision does not need to be final}. After they completed the decision-making task, they were asked to fill out a post-survey.

\paragraph{Interview \textit{(15 minutes)}.} After all three tasks, we conducted a semi-structured interview with the participants on their experience. We asked the participants to compare the three conditions in terms of the overall experience, in specific stages (i.e., establish a comprehensive understanding of the domain/the participant's situation, discover a diverse range of information, and manage the information found), and the final decision. We also asked \sysname{}-specific questions on the system features that helped information discovery, management, and making a final decision, the strengths and weaknesses of having multiple agents, and the potential use of \sysname{} in different types of decisions.

\subsubsection{Measures}
To observe the participants' decision-making processes and the outcomes, we collected interaction logs, surveys, and interview answers from the participants. 

\paragraph{Quantitative measures}
To understand the information space explored by the participants, we collected and counted the number of logs that indicated a search behavior, namely the user messages in \sysname{} and \multiagent{} and the search terms and clicking a new webpage in \web{}. We also collected the number of saving actions to the preference space, and counted them in terms of criterion and option.

\paragraph{Survey measures}
The post-task survey consisted of 7-point Likert-scale questions (1--strongly disagree, 7--strongly agree) measuring the effectiveness of the interface (broad information space, effective discovery and management of information, preference elicitation, quality of final decision), confidence in the decision~\cite{collier2012conflict}, satisfaction~\cite{lewis1991psychometric}, and NASA-TLX~\cite{HART1988139} for measuring workload. 

\paragraph{Analysis} 
For both quantitative and self-reported survey measures, we used Friedman tests with pairwise Wilcoxon signed-rank post-hoc analyses, applying Holm–Bonferroni correction for multiple comparisons. For the interviews, we first transcribed them automatically and used parts that helped interpret our statistical results. For analyzing RQ4: dominant strategies in using \sysname{}, we extracted the chat logs and two authors independently open-coded recurring usage patterns across participants. We then used interview transcripts to interpret participants' motivations behind these behaviors and iteratively refined the strategy categories through discussion, resulting in four dominant strategies. 

\subsection{Results}

\textbf{Summary}: In comparing \sysname{} to \web{}, \sysname{} successfully supported the discovery of a broad information space with lower burden. Participants also reported that an easier organization and structuring of relevant information was possible, leading to a significantly better understanding of their situation. While there was no significant difference in the perceived quality of the final decision, participants in \sysname{} were more satisfied with the process and confident in the decision.

In comparing \sysname{} to \multiagent{}, the exploration of information in the domain was significantly broader in \sysname{}. While there was no significant difference in discovering and managing relevant information, participants viewed the decision in \sysname{} significantly higher quality compared to \multiagent{}, with a tendency toward higher confidence in the decision.

We additionally report on four major strategies participants used for utilizing the multi-agents in \sysname{} to inform the design of a multi-agent conversational system.

\begin{figure*}[htb!]
    \centering
    \includegraphics[trim=0cm 0cm 0cm 0cm, clip=true, width=0.8\textwidth]{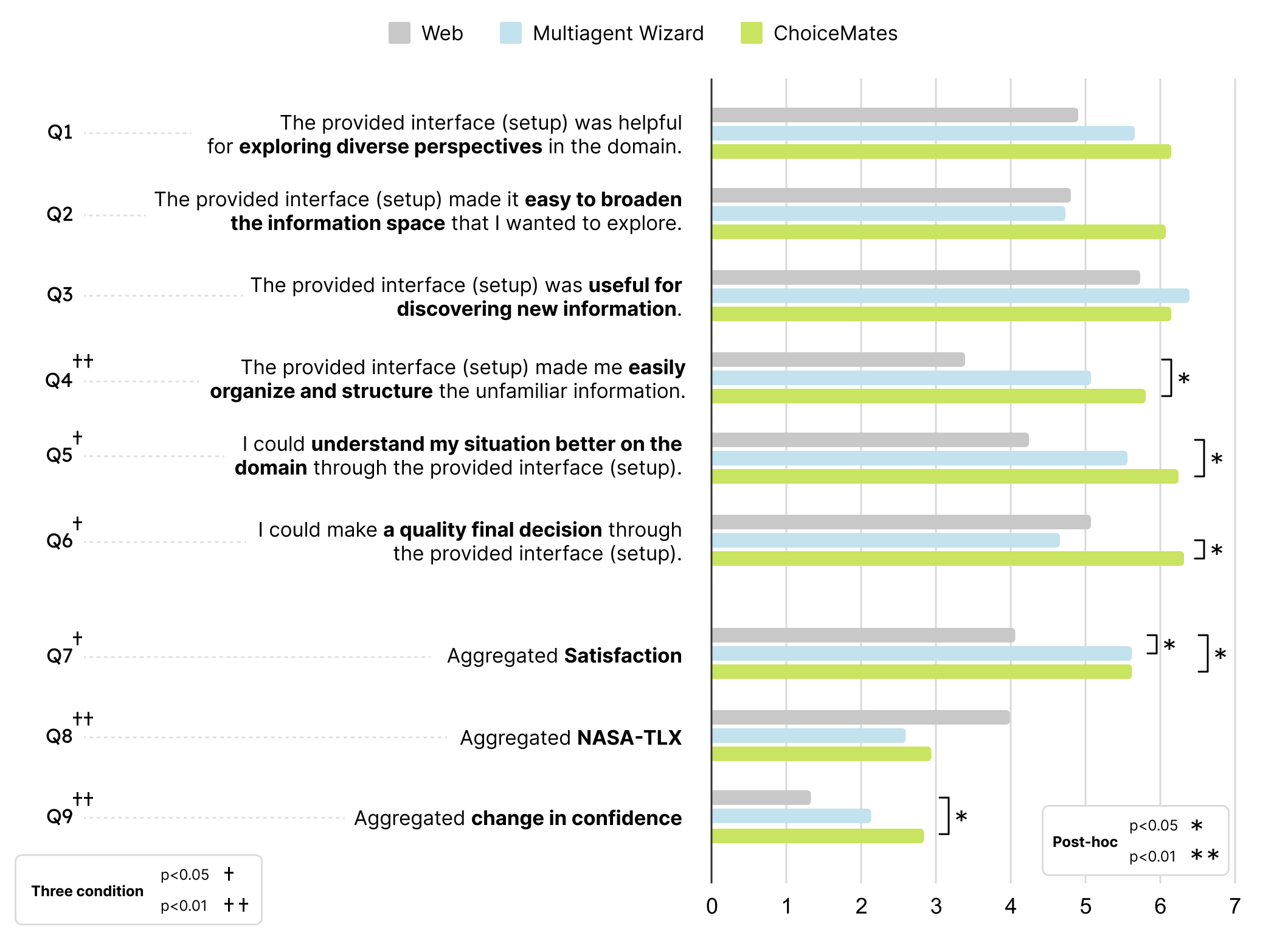}
    \caption{Analysis of the survey results.} 
    \Description{A bar graph that evaluates three user interfaces—Web, Multigent Wizard, and ChoiceMates—across nine questions related to usability, such as ease of exploring diverse perspectives and making quality decisions. Symbols denote statistical significance, with aggregated measures of satisfaction, NASA-TLX, and change in confidence at the bottom. The graph uses a seven-point scale, and statistical significance is indicated with asterisks.}
    \label{fig:survey}
\end{figure*}

\begin{figure*}[htb!]
\captionsetup{
  width=0.75\linewidth,
  justification=centering
}
    \centering
    \includegraphics[trim=0cm 0cm 0cm 0cm, clip=true, width=0.6\textwidth]{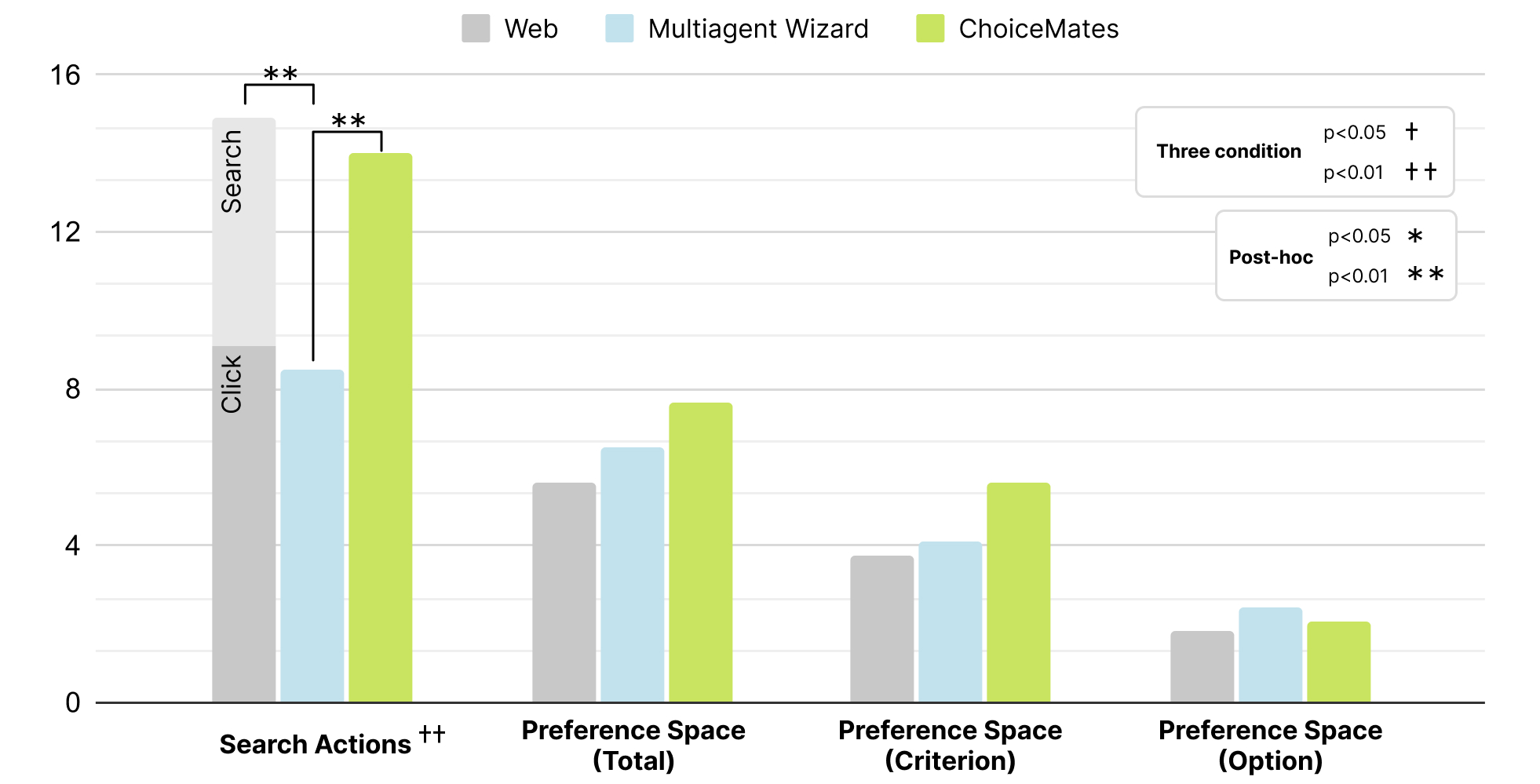}
    \caption{Analysis of the low-level actions. For search actions, we counted the number of search and click actions for \web{}, and the number of user messages for \multiagent{} and \sysname{}.} 
    \Description{A bar chart that compares the number of search and click actions in three interfaces: Web, Multiagent Wizard, and ChoiceMates. It highlights significant differences in 'Search' where Multigent Wizard exceeds others. The chart also shows 'Preference Space' actions, divided into 'Total,' 'Criterion,' and 'Option,' with notations for statistical significance.}
    \label{fig:quant}
\end{figure*}

\subsubsection{RQ1: How does \sysname{} help users explore a broad information space in the domain?}
 
With \sysname{}, participants indicated that they could explore a broad space of information, comparable to \web{}, and a broader space of information compared to \multiagent{}. 
While the web condition was most familiar to the participants and the search actions were easier to perform, there was no significant difference in the search action between \web{} and \sysname{} (Fig. \ref{fig:quant}; \web{}: M=14.917 (SD=6.067) / \multiagent{}: M=8.500 (SD= 2.203); \sysname{}: M=14.000 (SD=4.390); S=11.783, p=0.0028; \web{}-\sysname{}, p>0.05). This suggests that \sysname{} did not add interaction friction for exploration, as participants engaged in a comparable amount of exploratory actions as \web{} despite the learning curve.
In \multiagent{}, participants performed the least number of search actions significantly (\web{}-\multiagent{}, p<0.05; \sysname{}-\multiagent{}, p<0.05). This suggests that \sysname{} encouraged more active exploration than \multiagent{}, as reflected in the higher number of search actions participants performed.

When asked during the interview, ``In which condition are you able to discover a diverse range of information?'' seven participants chose \sysname{}, four opted for the \web{}, and one selected \multiagent{} condition. P9 described \multiagent{} as ``I merely followed through the agent's guidance, where I answered the agents' questions, received candidate options, and selected the one recommended among the options''. They commented that \multiagent{} helped in saving time for the decision, but as the process felt like an automated process they did not realize the need for expanding their information space. On the other hand, P9 mentioned that an agent constantly advocating for the criteria \criterion{portability} in \sysname{} led them to realize its importance and ask the question ``Is it portable?'' to the candidates to uncover more information related to portability.


\subsubsection{RQ2: How does \sysname{} support the discovery and management of relevant information to the user's context?}
Our survey results show that \sysname{} successfully supported organizing and structuring the information found and helped participants gain a better understanding of their situations to identify the relevant information compared to \web{}. 


Participants answered that \sysname{} let them easily organize and structure the unfamiliar information significantly compared to the \web{} condition (Fig. \ref{fig:survey}-Q4; \web{}: M=3.417 (SD=2.021) / \multiagent{}: M=5.083 (SD=1.881) / \sysname{}: M=5.833 (SD=1.193); S=11.400, p=0.0033; \sysname{}-\web{}, p<0.05). There was no significant difference between the \web{} and \multiagent{} (\multiagent{}-\web{}, p>0.05). While there were no significant differences, P1 noted that although \multiagent{} organizes information into bullet points, the delivery of text in a linear chat format without management support is inconvenient when you need to go back later in the chat.

The survey results also revealed that \sysname{} were more effective than the \web{} in assisting participants with understanding personal situations such as preference (Fig. \ref{fig:survey}-Q5; \web{}: M=4.250 (SD=1.960) / \multiagent{}: M=5.583 (SD=1.240) / \sysname{}: M=6.250 (SD=1.485); S=7.946, p=0.0188; \sysname{}-\web{}, p<0.05). There was also a tendency for \multiagent{} more effectively assisting situation understanding compared to \web{}, although not significant (\multiagent{}-\web{}, p>0.05).
In response to the interview question, ``Which interface helped you establish a comprehensive understanding of your situation?'', six participants mentioned \sysname{}, while the other six chose \multiagent{}. Regarding this, P10 mentioned that the basic questions provided by \multiagent{} in the initial stage such as ``What's your budget?'' helped them think more about their situation. P9 reported that in \sysname{}, the interactions with single or multiple agents giving a variety of options and criteria along with the conversation history that automatically logs this information, were beneficial for understanding their context, even within unfamiliar domains. On \web{}, there were no specific supports that were specifically designed to better elicit the user's context or preferences.


\subsubsection{RQ3: How does the user perceive their final decision with ~\sysname{}?}

There was a significant difference regarding the perceived quality of final decisions made between \sysname{} and \multiagent{} (Fig. \ref{fig:survey}-Q6; \web{}: M=5.083 (SD=1.676) / \multiagent{}: M=4.667 (SD=1.670) / \sysname{}: M=6.333 (SD=0.778); S=6.950, p=0.0310; \sysname{}-\multiagent{}, p<0.05). The decision confidence was descriptively higher in \sysname{} compared to \multiagent{}, although not significant (Fig. \ref{fig:survey}-Q9; \web{}: M=1.333 (SD=2.025) / \multiagent{}: M=2.139 (SD=1.141) / \sysname{}: M=2.861 (SD=1.573); S=9.957, p=0.0069; \sysname{}-\multiagent{}, p>0.05). P3 attributed their higher confidence to the interaction style: ``[\multiagent{}] felt like shopping---the store manager asking me what I want---whereas [\sysname{}] felt like a round table of experts.''


In comparing the decision between \sysname{} and \web{}, participants viewed the decision-making process as more confident (\sysname{}-\web{}, p<0.05) and satisfactory in \sysname{} (Fig. \ref{fig:survey}-Q7; \web{}: M=4.056 (SD=1.601) / \multiagent{}: M=5.639 (SD=0.979) / \sysname{}: M=5.639 (SD=0.893); S=8.711, p=0.0128; \sysname{}-\web{}, p<0.05). 

There were no significant differences in the total amount of saved options and criteria in the preference space between the conditions (Fig. \ref{fig:quant}; \web{}: M=5.583 (SD=2.610) / \multiagent{}: M=6.500 (SD=1.168) / \sysname{}: M=7.667 (SD=2.462); S=2.978, p>0.05). However, participants still perceived the process as more satisfactory and reported higher confidence during the process than \web{}, and they perceived the final decision as higher quality than \multiagent{}, with \sysname{} having the highest mean across all three factors. 
Connecting to the findings of RQ1, this suggests that for \multiagent{} participants might have perceived the process as satisfactory due to the agent taking the initiative. However, they still found that the final confidence and quality of the decision not as high as \sysname{}, as P1 stated ``I ended up choosing a product that I am not likely to buy, without considering my circumstances well enough.'' 
In \web{}, on the other hand, participants saw the final decision as of substantial quality but the process not as satisfactory. This suggests that \sysname{} can alleviate the burden of unfamiliar decision-making with \web{} while accomplishing a similar decision outcome.

\subsubsection{RQ4: How does the user utilize \sysname{} in the decision-making process?}

We observed four dominant strategies in how participants utilized \sysname{} for different needs in the decision-making process, and illustrate an end-to-end process with examples in Appendix \ref{appendix:p9-example}.

\paragraph{Talking to all agents to elicit relevant information in the domain.} When participants did not have any clue about the domain or have yet established any preferences in the domain, they chose to talk to all agents, so that they not only rely on \sysname{}'s response logic for more relevant agents or responses but also in understanding the perspectives in the domain through observing agent-agent conversations that occasionally appeared. Many of such conversations included the user's situation or preferences, such as P2 (climbing shoes)'s utterance ``Ok. then let's focus on the fact that I am a beginner.'' Criteria mentioned in the agents' responses were often saved in the preference space.

\paragraph{Tagging multiple agents to understand the domain and perspectives effectively.} After the participants had a decent understanding of the domain, they tagged multiple agents at once to quickly retrieve information and compare them.
When the participants successfully identified a few sufficient options, they asked the agents to ``debate'' or ``tell me more'' to elicit each agent's perspectives.
This also reflected the mental space of how many options participants had in mind, where P1's (climbing shoes) tagged number of agents went from 4, 3, 4, and 3 respectively throughout the process.

\paragraph{Conversing with one highly relevant agent.} When participants related themselves to a certain agent, they chose to converse further to inquire about all information and opinions the agent can provide. For example, P10 (fabric cleaner) conversed with Jordan (AlwaysLux EasyLint Professional Sweater Shaver) 4 out of 15 turns throughout the conversation, where the conversation not only contained specific questions (``Are there any color options?'') or elicitation of preferences (``I am leaning towards your device.''), but also asked for the experience behind it (``How long have you used it?''), expecting the answer from Jordan's profile.

\paragraph{Calling more agents to expand the information space.} When the agents in the space did not adequately reflect the participants' preferences, participants called in more agents into the space by asking ``any other agents''. This reflected the participant's desire to explore more options in the space, but also to double-check if they had considered the existing options sufficiently when they had the toggle button on for \sysname{} to take account their preferences. 



\subsubsection{Hallucination Analysis}

\begin{table*}[t]
\captionsetup{
  width=0.75\linewidth,
  justification=centering
}
\begin{tabular}{r|cc|c|c}
\hline
\multicolumn{1}{l|}{} & \multicolumn{2}{c|}{Factual Inaccuracy}                                                                                                                                   & \multirow{2}{*}{\begin{tabular}[c]{@{}c@{}}Irrelevant\\ Response\end{tabular}} & \multirow{2}{*}{\begin{tabular}[c]{@{}c@{}}Self-\\ Contradiction\end{tabular}} \\ \cline{2-3}
\multicolumn{1}{l|}{} & \multicolumn{1}{l}{\begin{tabular}[c]{@{}l@{}}Objective\\ Information\end{tabular}} & \multicolumn{1}{l|}{\begin{tabular}[c]{@{}l@{}}Subjective\\ Information\end{tabular}} &                                                                                &                                                                                \\ \hline
Total \#              & 439                                                                                & 981                                                                                  & 439                                                                            & 439                                                                            \\
Hallucination \#      & 8                                                                                  & 28                                                                                   & 21                                                                             & 5                                                                              \\ \hline
Hallucination \%      & 1.82\%                                                                             & 2.85\%                                                                               & 4.78\%                                                                         & 1.14\%                                                                         \\ \hline
\end{tabular}
\caption{Result of the hallucination analysis. Objective information refers to whether there was no incorrect factual information in the message, and subjective information refers to any parts of the message that reveal opinions through criteria about an option.}
\Description{This table consists of four rows and five columns, where the rows describe the type of hallucination analysis, total #, hallucination #, and hallucination \%. The columns consist of factual inaccuracy (objective information and subject information below), irrelevant response, and self-contradiction.}
\label{tab:hallucination-analysis}
\end{table*}

We conducted a small-scale hallucination analysis to understand the reliability of \sysname{}, as even a small hallucination could largely affect the user experience in unfamiliar decisions. We used the 439 agent conversations collected from the 12 sessions in the user study, and the three definitions of hallucination in Zhang et al.'s work \cite{zhang2023siren}, namely factual inaccuracy, irrelevant response, and self-contradiction. For each of the agent's messages, two of the authors individually coded 30\% of the messages, which resulted in a 95\% agreement rate. Then, they discussed to reach a consensus on the disagreements and individually coded the remaining messages. We describe the detailed criteria and the results (Table \ref{tab:hallucination-analysis}) for each definition below.

\paragraph{Factual Inaccuracy}
For factual inaccuracy, we first measured \textit{objective information} and \textit{subjective information} separately. Objective information refers to whether a message provides fact-related information (e.g., this vacuum cleaner is battery-operated), and subjective information refers to any parts of the message that reveal opinions through criteria about an option (e.g., The \%{battery life} is impressive for the \&{Roborock S6}). The total count of objective information was 439, and subjective information was 981. 

For objective information, we verified whether the information was correct by searching for information on the web. For subjective information, we marked the information as accurate if there existed an opinion on the first page of Google Search that described the option with the particular criteria (e.g., Roborock S6 good battery life), to ensure that the opinions come from an existing source. Since the agent's responses are generated by Google search results, we defined the correctness of subjective information and whether the information could be found in an actual Google search, not the strict correctness of the information.
The result yielded 1.82\% objective inaccuracy and 2.85\% subjective inaccuracy, where we saw a trend of once there was an inaccuracy in a piece of information, it continued throughout the conversation.

\paragraph{Irrelevant Response}
For irrelevant response, we looked for agent responses that did not reply in connection to the previous user message. The result yielded 4.78\% irrelevant response. The irrelevant response mostly consisted of the agents responding incorrectly to short utterances (e.g., irrelevant response from the agents when the user responded "250" for the question "What's your shoe size?"), or when the agents proactively spoke up even if they were not mentioned or their profile was irrelevant.

\paragraph{Self-contradiction}
For self-contradiction, we examined the consistency of agent responses by looking at the entire conversation and identifying parts where the agent did not adhere to its profile (i.e., valued criteria, option, and descriptor). The result yielded 1.14\% self-contradiction, which was 5 out of 439 messages total.

\section{Discussion}
In this section, we discuss the benefits of having agents as the basic unit of interaction and argue for the design of a more controllable and collaborative multi-agent system. We then discuss how \sysname{} may be utilized for other types of decisions and the impact of hallucination and potential information validation. We conclude with limitations and suggest possible future work.

\subsection{Agents as Interactables}
In \sysname{}, an \emph{agent} is the basic unit of the conversational interaction consisting of a descriptor, valued criteria, and a valued option, intended to help users identify the link between criteria and options. Our design for agents came from the first design goal, where we aimed to present unique experiences in the domain. 

By having agents as an information unit, the information that an agent contains becomes multifaceted: it could be factual information, unique experiences, an explanation of domain knowledge, or even questions that help the users to reflect on their situation. The exploration process can also become more flexible, where the user can ask and gain a wide variety of information solely by interacting with an agent to fulfill their dynamic inquiries.
Such a design complements a major limitation of the web in supporting information search, where usually one type of information is displayed within a single page, restricting the users in satisfying their information needs \cite{varadarajan2008beyond}. 

Moreover, in the multi-agent system, each agent embeds a certain kind of identity that is distinct from one another. With multiple agents each with its own identity, users can explore the information space more effectively and efficiently by using agents as a unit, with anchored profiles as memorable and engaging units of interaction.
We observed a similar trend where participants were more engaged in the unfamiliar decision-making process with \sysname{}. This led to some participants more eagerly describing their situation and preferences when they found the agents relatable, in return more precisely understanding their situation after an in-depth conversation. 
In unfamiliar decision-making where users can easily be overwhelmed by the abundance of information, multiple agents that are more memorable and engaging could motivate the users to actively explore the domain and secure better retention of the information found with less burden.


\subsection{Design of More Controllable and Collaborative Multi-agent System}

The key difference between \sysname{} and \multiagent{}, a commercialized multi-agent system, was that the users in \sysname{} had more agency in the decision-making process, while \multiagent{} guided the process for them. Our evaluation showed that participants in \sysname{} had significantly higher perceived decision quality on their decisions compared to participants in \multiagent{}. 
This may suggest that for unfamiliar decision domains, a multi-agent system with more agency can trigger a deeper reflection of participants' situations, as P11 mentioned, ``\sysname{}'s suggested options having links to valued criteria and options had me think more deeply about my priorities in the domain''. This echoes existing AI-assisted decision-making papers that found that providing second opinions or building upon users' decision flows led to better decisions and mitigated over-reliance \cite{lu2024does, reicherts2025ai}, suggesting the importance of agency in multi-agent interfaces.


Another aspect that \sysname{} brought in distinction to automated multi-agent systems is the sense of collaboration. While not quantitatively measured, many participants described the agents as if they were friends, indicating that ``hearing from multiple friends with experience was more helpful than from a single expert'' (P12). This perceived collegiality reduced the pressure to accept the first suggestion and made trade-offs more visible through contrastive viewpoints while the agents are highlighting independent opinions. At the same time, this sense of collaboration in a social setting could provide unexpected persuasion from agents being too human-like, highlighting the need for safeguards of collaboration while harnessing its benefits.

\subsection{\sysname{} for Other Types of Decisions} \label{discussion-3}

How generalizable is the multi-agent interaction to other types of decision-making tasks? 
\sysname{} is designed for decisions in unfamiliar domain, thus we put more emphasis on helping users easily understand diverse perspectives of the domain compared to familiar decisions. The multi-agent system shows advantages in discovering, managing, and comparing diverse opinions and experiences more easily, which was a pain point for unfamiliar decisions. This may suggest that \sysname{} could be utilized for more subjective decisions involving highly contrasting opinions (e.g., deciding who to vote in the next presidential election), with improved reasoning ability of State-Of-The-Art models such as OpenAI o1 \footnote{https://openai.com/index/learning-to-reason-with-llms/} for a more logically grounded delivery of opinions.

For high-stakes unfamiliar decisions, \sysname{} may be used for early stage exploration of the domain, in its ability to provide varying perspectives and important criteria in the domain. However, the need for simultaneous exploration of domain knowledge and personal preference may not apply throughout the decision-making process for high-stakes decision, careful examination of potential options and their risks is important for such decisions \cite{kunreuther2002high}. Additional supports to examine consequences and risks need to be incorporated, complementing the multi-agent approach \cite{gathani2025whatif}.

On the other hand, \sysname{} as a conversational interface inherently possesses several limitations. Information is primarily text-based and can easily become very cluttered, especially when there are multiple instances of it. Thus, the current interactions may not be suitable for decisions heavily involving multimodal data (e.g., when aesthetics becomes essential), and for more complex decisions with hierarchies (e.g., choosing a supplier) or multiple steps (e.g., trip planning). If such decisions are unfamiliar to the user, preventing users from being overwhelmed should be the main priority. In such cases, the agent unit for presenting information could remain, but the conversational and the preference construction aspect could reflect the characteristic of the decision domain to prevent information clutter.

\subsection{Impact of Hallucination and Information Validation}
We observed a small proportion of hallucination behaviors in \sysname{} with the hallucination analysis. As unfamiliar decision-making is a highly information-heavy task, even a little hallucination could influence the final decision or the trust towards certain agents.

Thus, interventions must be designed to help the user validate the information. While \sysname{} provides a hyperlink on options to let the user verify incorrect information or agent profiles to check consistency, future interactions could incorporate more proactive approaches to help users spot and mitigate misinformation. For example, hallucinations could be detected as a pipeline through detection algorithms (e.g., the HaluEval 2.0 benchmark \cite{li2024dawn}), and the system or another agent can come up with strategies to not only visually alert the user about hallucinations but also guide users to be able to evaluate hallucination moving forward. Furthermore, we believe that state-of-the-art technologies such as SearchGPT \cite{openai2024searchgpt} or Perplexity \cite{perplexityai2024perplexityai} would eventually minimize incorrect information or inconsistent behaviors, allowing users to fully utilize the agents to discover and understand deeper and quality information.

We also suggest that the opinions of LLM agents in such a task must be factually grounded. The subjective information provided by the agents in \sysname{} was a mix of factual information and opinions (e.g., ``The Roomba i7+ that I use has a pretty decent battery life. It can run for about 75 minutes before it needs to recharge.'') where opinions explain how valued criteria maps to options chosen by each agent. While more human-like conversations could highly engage the user in using \sysname{}, if the LLM agents provide information without factual evidence (e.g., ``This TV is the best since it gives me a good vibe.''), such interactions may foster reliance on unfounded assertions, ultimately impacting decision quality.

\subsection{Limitations and Future Work}
While \sysname{} was designed and evaluated carefully, there are some inherent limitations that this work carries, including the design of agents, potential bias in deployment, and evaluation setup.

\paragraph{Design of agents for information seeking.}
Unlike web search where users can directly inspect original sources, \sysname{} intermediates information through agent representations. Even when the information is grounded in retrieved web content, this mediation can obscure what evidence supports an opinion, making it hard for users to judge credibility. In addition, the agent personas in \sysname{} may not reflect real-world individuals or authentic lived experiences, which may trigger false beliefs when conversations become more personal and subjective. Paradoxically, these same social cues may also encourage over-anthropomorphism and over-trust, especially during subjective or emotional exchanges.
To reduce the risk of false beliefs and over-anthropomorphization of agents, future work could provide claim-level provenance (e.g., citations linked to specific statements), calibrate persona claims using external evidence, or offer user controls to soften or disable persona framing (e.g., switching to a neutral, evidence-first style when the system is uncertain).

\paragraph{Potential risks in deployment.}
Because \sysname{} can shape which options and criteria users consider, it could amplify biases in retrieved sources or be misused to steer users toward particular outcomes, for instance, through sponsored agents. These risks are especially salient for potential commercial deployment. Future work should study mitigation strategies such as disclosure of sponsorship or incentives, audits for systematic bias, and interaction designs that preserve user agency, such as through counter-arguments and imposing viewpoint diversity constraints.

\paragraph{Study settings.} Our study was carried out as a lab study, where participants were given a pre-assigned decision domain and a relatively short time frame (20 minutes) to make the decision. This was rather limiting to observe each step of the decision-making process---information seeking, understanding, and decision-making---and the aftereffects of the decision.  
The relatively small sample size (n=12 per condition) also limits statistical power and may introduce random variability in the results; non-significant findings should therefore be interpreted as inconclusive. Future work should adopt longer-term, larger-scale, in-situ field deployments to more robustly evaluate the effects of \sysname{}.

\section{Conclusion}

In this work, we propose \sysname{}, a multi-agent conversational system that supports unfamiliar decision-making. Our study comparing \sysname{} to a web search condition and a multi-agent framework condition reveals that \sysname{} can support a broader exploration of information in the decision space and a higher-quality decision compared to \multiagent{}, and can better organize and structure the information found, leading to a more satisfied process and confident final decision compared to \web{}. We describe different strategies used to orchestrate the agents in \sysname{} for unfamiliar decision-making, then discuss user-side implications of a multi-agent conversational system for information discovery and decision-making support.


\section{GenAI Usage Disclosure}

Other than the large language model used to develop the core functionalities of \sysname{}, generative AI was not used in any part of the research process other than writing assistance. During the writing process, the authors used generative AI to proofread and detect minor grammar issues and misprints.

\begin{acks}
This work was supported by the National Research Foundation of Korea (NRF) grant funded by the Korea government (Ministry of Science and ICT) (No.RS-2025-00557726), and Institute of Information \& communications Technology Planning \& Evaluation (IITP) grant funded by the Korea government (MSIT) (No.RS-2024-00443251, Accurate and Safe Multimodal, Multilingual Personalized AI Tutors). This work was also supported by the National Research Foundation of Korea (NRF) grants funded by the Korea government (MSIT) (Nos. RS-2024-00353125 and 2022R1C1C1003123).
The first author would like to thank the co-authors for their contribution to this exciting idea. Furthermore, we thank all of our study participants, reviewers, and the members of KIXLAB for their insightful discussions and constructive feedback.
\end{acks}


\bibliographystyle{ACM-Reference-Format}
\bibliography{main}

@article{jeon2013value,
  title={The value of social search: Seeking collective personal experience in social Q\&A},
  author={Jeon, Grace YoungJoo and Rieh, Soo Young},
  journal={Proceedings of the American Society for Information Science and Technology},
  volume={50},
  number={1},
  pages={1--10},
  year={2013},
  publisher={Wiley Online Library}
}

@article{li2024dawn,
  title={The dawn after the dark: An empirical study on factuality hallucination in large language models},
  author={Li, Junyi and Chen, Jie and Ren, Ruiyang and Cheng, Xiaoxue and Zhao, Wayne Xin and Nie, Jian-Yun and Wen, Ji-Rong},
  journal={arXiv preprint arXiv:2401.03205},
  year={2024}
}

@misc{openai2024searchgpt,
  author       = {{OpenAI}},
  title        = {SearchGPT Prototype},
  year         = {2024},
  url          = {https://openai.com/index/searchgpt-prototype/},
  note         = {Accessed: 2024-09-11}
}

@misc{perplexityai2024perplexityai,
  author       = {{Perplexity AI}},
  title        = {Perplexity AI},
  year         = {2024},
  url          = {https://www.perplexity.ai/},
  note         = {Accessed: 2024-09-11}
}

@article{yuan2024mora,
  title={Mora: Enabling generalist video generation via a multi-agent framework},
  author={Yuan, Zhengqing and Chen, Ruoxi and Li, Zhaoxu and Jia, Haolong and He, Lifang and Wang, Chi and Sun, Lichao},
  journal={arXiv preprint arXiv:2403.13248},
  year={2024}
}

@article{wang2024learning,
      title={Learning to Break: Knowledge-Enhanced Reasoning in Multi-Agent Debate System}, 
      author={Haotian Wang and Xiyuan Du and Weijiang Yu and Qianglong Chen and Kun Zhu and Zheng Chu and Lian Yan and Yi Guan},
      year={2024},
      journal={arXiv preprint arXiv:2312.04854},
}

@article{li2023traineragent,
  title={TrainerAgent: Customizable and Efficient Model Training through LLM-Powered Multi-Agent System},
  author={Li, Haoyuan and Jiang, Hao and Zhang, Tianke and Yu, Zhelun and Yin, Aoxiong and Cheng, Hao and Fu, Siming and Zhang, Yuhao and He, Wanggui},
  journal={arXiv preprint arXiv:2311.06622},
  year={2023}
}

@inproceedings{ccelen2024design,
author = {\c{C}elen, Ata and Han, Guo and Schindler, Konrad and Van Gool, Luc and Armeni, Iro and Obukhov, Anton and Wang, Xi},
title = {I-Design: Personalized LLM Interior Designer},
year = {2025},
isbn = {978-3-031-92386-9},
publisher = {Springer-Verlag},
address = {Berlin, Heidelberg},
url = {https://doi.org/10.1007/978-3-031-92387-6_17},
doi = {10.1007/978-3-031-92387-6_17},
booktitle = {Computer Vision – ECCV 2024 Workshops: Milan, Italy, September 29–October 4, 2024, Proceedings, Part II},
pages = {217–234},
numpages = {18},
location = {Milan, Italy}
}

@article{inoue2024drugagent,
  title={DrugAgent: Explainable Drug Repurposing Agent with Large Language Model-based Reasoning},
  author={Inoue, Yoshitaka and Song, Tianci and Fu, Tianfan},
  journal={arXiv preprint arXiv:2408.13378},
  year={2024}
}

@inproceedings{fan2024contextcam,
  title={ContextCam: Bridging Context Awareness with Creative Human-AI Image Co-Creation},
  author={Fan, Xianzhe and Wu, Zihan and Yu, Chun and Rao, Fenggui and Shi, Weinan and Tu, Teng},
  booktitle={Proceedings of the CHI Conference on Human Factors in Computing Systems},
  pages={1--17},
  year={2024}
}

@inproceedings{schneider2023investigating,
  title={Investigating conversational search behavior for domain exploration},
  author={Schneider, Phillip and Afzal, Anum and Vladika, Juraj and Braun, Daniel and Matthes, Florian},
  booktitle={European Conference on Information Retrieval},
  pages={608--616},
  year={2023},
  organization={Springer}
}

@article{liao2020conversational,
  title={Conversational interfaces for information search},
  author={Liao, Q Vera and Geyer, Werner and Muller, Michael and Khazaen, Yasaman},
  journal={Understanding and Improving Information Search: A Cognitive Approach},
  pages={267--287},
  year={2020},
  publisher={Springer}
}

@inproceedings{agrawal2015whither,
author = {Agrawal, Rakesh and Golshan, Behzad and Papalexakis, Evangelos},
title = {Whither Social Networks for Web Search?},
year = {2015},
isbn = {9781450336642},
publisher = {Association for Computing Machinery},
address = {New York, NY, USA},
url = {https://doi.org/10.1145/2783258.2788571},
doi = {10.1145/2783258.2788571},
booktitle = {Proceedings of the 21th ACM SIGKDD International Conference on Knowledge Discovery and Data Mining},
pages = {1661–1670},
numpages = {10},
location = {Sydney, NSW, Australia},
series = {KDD '15}
}

@article{liu2024lost,
    title = "Lost in the Middle: How Language Models Use Long Contexts",
    author = "Liu, Nelson F.  and
      Lin, Kevin  and
      Hewitt, John  and
      Paranjape, Ashwin  and
      Bevilacqua, Michele  and
      Petroni, Fabio  and
      Liang, Percy",
    journal = "Transactions of the Association for Computational Linguistics",
    volume = "12",
    year = "2024",
    address = "Cambridge, MA",
    publisher = "MIT Press",
    url = "https://aclanthology.org/2024.tacl-1.9",
    doi = "10.1162/tacl_a_00638",
    pages = "157--173",
}

@article{zhang2023siren,
  title={Siren's song in the AI ocean: a survey on hallucination in large language models},
  author={Zhang, Yue and Li, Yafu and Cui, Leyang and Cai, Deng and Liu, Lemao and Fu, Tingchen and Huang, Xinting and Zhao, Enbo and Zhang, Yu and Chen, Yulong and others},
  journal={arXiv preprint arXiv:2309.01219},
  year={2023}
}

@article{guo2022can,
  title={Can the amount of information and information presentation reduce choice overload? An empirical study of online hotel booking},
  author={Guo, Rui and Li, Hengyun},
  journal={Journal of Travel \& Tourism Marketing},
  volume={39},
  number={1},
  pages={87--108},
  year={2022},
  publisher={Taylor \& Francis}
}

@inproceedings{lunenburg2010decision,
  title={The decision making process.},
  author={Lunenburg, Fred C},
  booktitle={National Forum of Educational Administration \& Supervision Journal},
  volume={27},
  number={4},
  year={2010}
}

@inproceedings{knijnenburg2009adaptivePE,
author = {Knijnenburg, Bart P. and Willemsen, Martijn C.},
title = {Understanding the Effect of Adaptive Preference Elicitation Methods on User Satisfaction of a Recommender System},
year = {2009},
isbn = {9781605584355},
publisher = {Association for Computing Machinery},
address = {New York, NY, USA},
url = {https://doi.org/10.1145/1639714.1639793},
doi = {10.1145/1639714.1639793},
booktitle = {Proceedings of the Third ACM Conference on Recommender Systems},
pages = {381–384},
numpages = {4},
location = {New York, New York, USA},
series = {RecSys '09}
}

@article{liu2023selenite,
  title={Selenite: Scaffolding Online Sensemaking with Comprehensive Overviews Elicited from Large Language Models},
  author={Liu, Michael Xieyang and Wu, Tongshuang and Chen, Tianying and Li, Franklin Mingzhe and Kittur, Aniket and Myers, Brad A},
  journal={arXiv preprint arXiv:2310.02161},
  year={2023}
}

@ARTICLE{Jannach2021-rg,
  title     = "A Survey on Conversational Recommender Systems",
  author    = "Jannach, Dietmar and Manzoor, Ahtsham and Cai, Wanling and Chen,
               Li",
  journal   = "ACM Comput. Surv.",
  publisher = "Association for Computing Machinery",
  volume    =  54,
  number    =  5,
  pages     = "1--36",
  month     =  may,
  year      =  2021,
  url       = "https://doi.org/10.1145/3453154",
  address   = "New York, NY, USA",
  issn      = "0360-0300",
  doi       = "10.1145/3453154"
}

@ARTICLE{Warnestal2005-yv,
  title     = "User evaluation of a conversational recommender system",
  author    = "W{\"a}rnest{\aa}l, Pontus",
  journal   = "Proceedings of the 4th IJCAI Workshop on …",
  publisher = "csse.monash.edu.au",
  month     =  jan,
  year      =  2005,
  url       = "https://www.academia.edu/1893257/User_evaluation_of_a_conversational_recommender_system",
  language  = "en"
}

@ARTICLE{Wang2013-hd,
  title     = "Research Note: A Contingency Approach to Investigating the
               Effects of {User-System} Interaction Modes of Online Decision
               Aids",
  author    = "Wang, Weiquan and Benbasat, Izak",
  journal   = "Information Systems Research",
  publisher = "INFORMS",
  volume    =  24,
  number    =  3,
  pages     = "861--876",
  year      =  2013,
  url       = "http://www.jstor.org/stable/42004297",
  issn      = "1526-5536"
}

@INPROCEEDINGS{Liu2022-crystalline,
  title     = "Crystalline: Lowering the Cost for Developers to Collect and
               Organize Information for Decision Making",
  booktitle = "Proceedings of the 2022 {CHI} Conference on Human Factors in
               Computing Systems",
  author    = "Liu, Michael Xieyang and Kittur, Aniket and Myers, Brad A",
  publisher = "Association for Computing Machinery",
  number    = "Article 68",
  pages     = "1--16",
  series    = "CHI '22",
  month     =  apr,
  year      =  2022,
  url       = "https://doi.org/10.1145/3491102.3501968",
  address   = "New York, NY, USA",
  location  = "New Orleans, LA, USA",
  isbn      = "9781450391573",
  doi       = "10.1145/3491102.3501968"
}

@INPROCEEDINGS{Liu2019-unakite,
  title     = "Unakite: Scaffolding Developers' {Decision-Making} Using the Web",
  booktitle = "Proceedings of the 32nd Annual {ACM} Symposium on User Interface
               Software and Technology",
  author    = "Liu, Michael Xieyang and Hsieh, Jane and Hahn, Nathan and Zhou,
               Angelina and Deng, Emily and Burley, Shaun and Taylor, Cynthia
               and Kittur, Aniket and Myers, Brad A",
  publisher = "Association for Computing Machinery",
  pages     = "67--80",
  series    = "UIST '19",
  month     =  oct,
  year      =  2019,
  url       = "https://doi.org/10.1145/3332165.3347908",
  address   = "New York, NY, USA",
  location  = "New Orleans, LA, USA",
  isbn      = "9781450368162",
  doi       = "10.1145/3332165.3347908"
}

@INPROCEEDINGS{Chang2020-mesh,
  title     = "Mesh: Scaffolding Comparison Tables for Online Decision Making",
  booktitle = "Proceedings of the 33rd Annual {ACM} Symposium on User Interface
               Software and Technology",
  author    = "Chang, Joseph Chee and Hahn, Nathan and Kittur, Aniket",
  publisher = "Association for Computing Machinery",
  pages     = "391--405",
  series    = "UIST '20",
  month     =  oct,
  year      =  2020,
  url       = "https://doi.org/10.1145/3379337.3415865",
  address   = "New York, NY, USA",
  location  = "Virtual Event, USA",
  isbn      = "9781450375146",
  doi       = "10.1145/3379337.3415865"
}

@article{jiang2023graphologue,
  title={Graphologue: Exploring Large Language Model Responses with Interactive Diagrams},
  author={Jiang, Peiling and Rayan, Jude and Dow, Steven and Xia, Haijun},
  journal={arXiv preprint arXiv:2305.11473},
  year={2023}
}

@article{wu2023autogen,
    title={AutoGen: Enabling Next-Gen LLM Applications via Multi-Agent Conversation Framework}, 
    author={Qingyun Wu and Gagan Bansal and Jieyu Zhang and Yiran Wu and Shaokun Zhang and Erkang Zhu and Beibin Li and Li Jiang and Xiaoyun Zhang and Chi Wang},
    year={2023},
    journal={arXiv preprint arXiv:2308.08155},
}

@ARTICLE{Bettman1980-phasechoiceprocess,
  title     = "Effects of Prior Knowledge and Experience and Phase of the
               Choice Process on Consumer Decision Processes: A Protocol
               Analysis",
  author    = "Bettman, James R and Park, C Whan",
  journal   = "J. Consum. Res.",
  publisher = "Oxford Academic",
  volume    =  7,
  number    =  3,
  pages     = "234--248",
  month     =  dec,
  year      =  1980,
  url       = "https://academic.oup.com/jcr/article-abstract/7/3/234/1820883",
  language  = "en",
  issn      = "0093-5301",
  doi       = "10.1086/208812"
}

@ARTICLE{Karimi2015-priorknowledgetostyle,
  title    = "The effect of prior knowledge and decision-making style on the
              online purchase decision-making process: A typology of consumer
              shopping behaviour",
  author   = "Karimi, Sahar and Papamichail, K Nadia and Holland, Christopher P",
  journal  = "Decis. Support Syst.",
  volume   =  77,
  pages    = "137--147",
  month    =  sep,
  year     =  2015,
  url      = "https://www.sciencedirect.com/science/article/pii/S0167923615001189",
  issn     = "0167-9236",
  doi      = "10.1016/j.dss.2015.06.004"
}

@inbook{Candello2016-designconvui,
author = {Candello, Heloisa and Pinhanez, Claudio},
year = {2016},
month = {10},
pages = {http://www.lbd.dcc.ufmg.br/bdbcomp/servlet/Trabalho?id=25402},
publisher = {BDBComp},
address = {},
title = {Designing Conversational Interfaces}
}

@INPROCEEDINGS{Radlinski2017-convsearchframework,
  title     = "A Theoretical Framework for Conversational Search",
  booktitle = "Proceedings of the 2017 Conference on Conference Human
               Information Interaction and Retrieval",
  author    = "Radlinski, Filip and Craswell, Nick",
  publisher = "Association for Computing Machinery",
  pages     = "117--126",
  series    = "CHIIR '17",
  month     =  mar,
  year      =  2017,
  url       = "https://doi.org/10.1145/3020165.3020183",
  address   = "New York, NY, USA",
  location  = "Oslo, Norway",
  isbn      = "9781450346771",
  doi       = "10.1145/3020165.3020183"
}

@article{malhotra1983individual,
  title={On'Individual differences in search behavior for a nondurable'},
  author={Malhotra, Naresh K},
  journal={Journal of Consumer Research},
  volume={10},
  number={1},
  pages={125--131},
  year={1983},
  publisher={JSTOR}
}

@article{moore1980individual,
  title={Individual differences in search behavior for a nondurable},
  author={Moore, William L and Lehmann, Donald R},
  journal={Journal of consumer research},
  volume={7},
  number={3},
  pages={296--307},
  year={1980},
  publisher={The University of Chicago Press}
}

@article{chen2013human,
author = {Chen, Li and deGemmis, Marco and Felfernig, Alexander and Lops, Pasquale and Ricci, Francesco and Semeraro, Giovanni},
year = {2013},
month = {10},
pages = {},
title = {Human Decision Making and Recommender Systems},
volume = {3},
journal = {ACM Transactions on Interactive Intelligent Systems},
doi = {10.1145/2533670.2533675}
}

@article{iyengar2000choice,
  title={When choice is demotivating: Can one desire too much of a good thing?},
  author={Iyengar, Sheena S and Lepper, Mark R},
  journal={Journal of personality and social psychology},
  volume={79},
  number={6},
  pages={995},
  year={2000},
  publisher={American Psychological Association}
}

@article{jameson2015human,
  title={Human decision making and recommender systems},
  author={Jameson, Anthony and Willemsen, Martijn C and Felfernig, Alexander and De Gemmis, Marco and Lops, Pasquale and Semeraro, Giovanni and Chen, Li},
  journal={Recommender systems handbook},
  pages={611--648},
  year={2015},
  publisher={Springer}
}

@article{thai2017too,
  title={Too many destinations to visit: Tourists’ dilemma?},
  author={Thai, Nguyen T and Yuksel, Ulku},
  journal={Annals of Tourism Research},
  volume={62},
  pages={38--53},
  year={2017},
  publisher={Elsevier}
}

@article{scheibehenne2010can,
  title={Can there ever be too many options? A meta-analytic review of choice overload},
  author={Scheibehenne, Benjamin and Greifeneder, Rainer and Todd, Peter M},
  journal={Journal of consumer research},
  volume={37},
  number={3},
  pages={409--425},
  year={2010},
  publisher={The University of Chicago Press}
}

@article{haubl2000consumer,
  title={Consumer decision making in online shopping environments: The effects of interactive decision aids},
  author={H{\"a}ubl, Gerald and Trifts, Valerie},
  journal={Marketing science},
  volume={19},
  number={1},
  pages={4--21},
  year={2000},
  publisher={INFORMS}
}

@inproceedings{Nourani2020TheRO,
  title={The Role of Domain Expertise in User Trust and the Impact of First Impressions with Intelligent Systems},
  author={Mahsan Nourani and Joanie T. King and Eric D. Ragan},
  booktitle={AAAI Conference on Human Computation \& Crowdsourcing},
  year={2020},
  url={https://api.semanticscholar.org/CorpusID:221186776}
}

@article{lops2011content,
  title={Content-based recommender systems: State of the art and trends},
  author={Lops, Pasquale and De Gemmis, Marco and Semeraro, Giovanni},
  journal={Recommender systems handbook},
  pages={73--105},
  year={2011},
  publisher={Springer}
}

@incollection{schafer2007collaborative,
  title={Collaborative filtering recommender systems},
  author={Schafer, J Ben and Frankowski, Dan and Herlocker, Jon and Sen, Shilad},
  booktitle={The adaptive web: methods and strategies of web personalization},
  pages={291--324},
  year={2007},
  publisher={Springer}
}

@inproceedings{collier2012conflict,
author = {Collier, Benjamin and Bear, Julia},
title = {Conflict, Criticism, or Confidence: An Empirical Examination of the Gender Gap in Wikipedia Contributions},
year = {2012},
isbn = {9781450310864},
publisher = {Association for Computing Machinery},
address = {New York, NY, USA},
url = {https://doi.org/10.1145/2145204.2145265},
doi = {10.1145/2145204.2145265},
booktitle = {Proceedings of the ACM 2012 Conference on Computer Supported Cooperative Work},
pages = {383–392},
numpages = {10},
location = {Seattle, Washington, USA},
series = {CSCW '12}
}

@article{lewis1991psychometric,
author = {Lewis, James R.},
title = {Psychometric Evaluation of an After-Scenario Questionnaire for Computer Usability Studies: The ASQ},
year = {1991},
issue_date = {Jan. 1991},
publisher = {Association for Computing Machinery},
address = {New York, NY, USA},
volume = {23},
number = {1},
issn = {0736-6906},
url = {https://doi.org/10.1145/122672.122692},
doi = {10.1145/122672.122692},
journal = {SIGCHI Bull.},
month = {jan},
pages = {78–81},
numpages = {4}
}

@incollection{HART1988139,
    title = {Development of NASA-TLX (Task Load Index): Results of Empirical and Theoretical Research},
    editor = {Peter A. Hancock and Najmedin Meshkati},
    series = {Advances in Psychology},
    publisher = {North-Holland},
    volume = {52},
    pages = {139-183},
    year = {1988},
    booktitle = {Human Mental Workload},
    issn = {0166-4115},
    doi = {https://doi.org/10.1016/S0166-4115(08)62386-9},
    url = {https://www.sciencedirect.com/science/article/pii/S0166411508623869},
    author = {Sandra G. Hart and Lowell E. Staveland}
}

@inproceedings{kitamura2002multiple,
author = {Kitamura, Yasuhiko and Tsujimoto, Hideki and Yamada, Teruhiro and Yamamoto, Taizo},
title = {Multiple Character-Agents Interface: An Information Integration Platform Where Multiple Agents and Human User Collaborate},
year = {2002},
isbn = {1581134800},
publisher = {Association for Computing Machinery},
address = {New York, NY, USA},
url = {https://doi.org/10.1145/544862.544925},
doi = {10.1145/544862.544925},
booktitle = {Proceedings of the First International Joint Conference on Autonomous Agents and Multiagent Systems: Part 2},
pages = {790–791},
numpages = {2},
location = {Bologna, Italy},
series = {AAMAS '02}
}

@article{kitamura2004web,
  title={Web information integration using multiple character agents},
  author={Kitamura, Yasuhiko},
  journal={Life-Like Characters: Tools, Affective Functions, and Applications},
  pages={295--315},
  year={2004},
  publisher={Springer}
}

@inproceedings{li2023leaders,
author = {Li, Shuo and Yuan, Xiang and Zhao, Xinyuan and Yang, Shirao},
title = {Leaders or Team-Mates: Exploring the Role-Based Relationship Between Multiple Intelligent Agents in Driving Scenarios: Research on the Role-Based Relationship Between Multiple Intelligent Agents in Driving Scenarios},
year = {2023},
isbn = {978-3-031-35677-3},
publisher = {Springer-Verlag},
address = {Berlin, Heidelberg},
url = {https://doi.org/10.1007/978-3-031-35678-0_9},
doi = {10.1007/978-3-031-35678-0_9},
pages = {144–165},
numpages = {22},
location = {Copenhagen, Denmark}
}

@inproceedings{park2022socialsimulacra,
author = {Park, Joon Sung and Popowski, Lindsay and Cai, Carrie and Morris, Meredith Ringel and Liang, Percy and Bernstein, Michael S.},
title = {Social Simulacra: Creating Populated Prototypes for Social Computing Systems},
year = {2022},
isbn = {9781450393201},
publisher = {Association for Computing Machinery},
address = {New York, NY, USA},
url = {https://doi.org/10.1145/3526113.3545616},
doi = {10.1145/3526113.3545616},
booktitle = {Proceedings of the 35th Annual ACM Symposium on User Interface Software and Technology},
articleno = {74},
numpages = {18},
location = {Bend, OR, USA},
series = {UIST '22}
}

@article{park2023generativeagents,
    author = {Park, Joon Sung and O'Brien, Joseph C. and Cai, Carrie J. and Morris, Meredith Ringel and Liang Percy and Bernstein, Michael S.},
    title = {Generative Agents: Interactive Simulacra of Human Behavior},
    journal={arXiv preprint arXiv:2304.03442},
    year = {2023}
}

@article{cox2023prompting,
      title={Prompting a Large Language Model to Generate Diverse Motivational Messages: A Comparison with Human-Written Messages}, 
      author={Cox, Samuel Rhys and Abdul, Ashraf and Ooi, Wei Tsang},
      journal={arXiv preprint arXiv:2308.13479},
      year={2023},
}

@inproceedings{
chan2023chateval,
title={ChatEval: Towards Better {LLM}-based Evaluators through Multi-Agent Debate},
author={Chi-Min Chan and Weize Chen and Yusheng Su and Jianxuan Yu and Wei Xue and Shanghang Zhang and Jie Fu and Zhiyuan Liu},
booktitle={The Twelfth International Conference on Learning Representations},
year={2024},
url={https://openreview.net/forum?id=FQepisCUWu}
}

@article{liu2023lost,
      title={Lost in the Middle: How Language Models Use Long Contexts}, 
      author={Liu, Nelson F. and Lin, Kevin and Hewitt, John and Paranjape, Ashwin and Bevilacqua, Michele and Petroni, Fabio and Liang, Percy},
      year={2023},
      journal={arXiv preprint arXiv:2307.03172},
}

@misc{langchain2023compression,
    title={LangChain Contextual Compression},
    year={2023},
    author={LangChain},
    url={https://python.langchain.com/docs/modules/data_connection/retrievers/contextual_compression/},
    note={Accessed: September 10, 2023}
}

@article{chen2023agentverse,
      title={AgentVerse: Facilitating Multi-Agent Collaboration and Exploring Emergent Behaviors in Agents}, 
      author={Chen, Weize and Su, Yusheng and Zuo, Jingwei and Yang, Cheng and Yuan, Chenfei and Qian, Chen and Chan, Chi-Min and Qin, Yujia and Lu, Yaxi and Xie, Ruobing and Liu, Zhiyuan and Sun, Maosong and Zhou, Jie},
      year={2023},
      journal={arXiv preprint arXiv:2308.10848},
}

@ARTICLE{zue2000cuiadvances,
  author={Zue, V.W. and Glass, J.R.},
  journal={Proceedings of the IEEE}, 
  title={Conversational interfaces: advances and challenges}, 
  year={2000},
  volume={88},
  number={8},
  pages={1166-1180},
  doi={10.1109/5.880078}}

@article{lewis2020retrieval,
  title={Retrieval-augmented generation for knowledge-intensive nlp tasks},
  author={Lewis, Patrick and Perez, Ethan and Piktus, Aleksandra and Petroni, Fabio and Karpukhin, Vladimir and Goyal, Naman and K{\"u}ttler, Heinrich and Lewis, Mike and Yih, Wen-tau and Rockt{\"a}schel, Tim and others},
  journal={Advances in Neural Information Processing Systems},
  volume={33},
  pages={9459--9474},
  year={2020}
}

@article{curcseu2015cognitive,
  title={Cognitive synergy in groups and group-to-individual transfer of decision-making competencies},
  author={Cur{\c{s}}eu, Petru L and Meslec, Nicoleta and Pluut, Helen and Lucas, Gerardus JM},
  journal={Frontiers in psychology},
  volume={6},
  pages={1375},
  year={2015},
  publisher={Frontiers Media SA}
}

@inproceedings{goertzel2017formal,
  title={A formal model of cognitive synergy},
  author={Goertzel, Ben},
  booktitle={Artificial General Intelligence: 10th International Conference, AGI 2017, Melbourne, VIC, Australia, August 15-18, 2017, Proceedings 10},
  pages={13--22},
  year={2017},
  organization={Springer}
}

@article{jiang2023communitybots,
  title={CommunityBots: Creating and Evaluating A Multi-Agent Chatbot Platform for Public Input Elicitation},
  author={Jiang, Zhiqiu and Rashik, Mashrur and Panchal, Kunjal and Jasim, Mahmood and Sarvghad, Ali and Riahi, Pari and DeWitt, Erica and Thurber, Fey and Mahyar, Narges},
  journal={Proceedings of the ACM on Human-Computer Interaction},
  volume={7},
  number={CSCW1},
  pages={1--32},
  year={2023},
  publisher={ACM New York, NY, USA}
}

@inproceedings{zolitschka2020novel,
  title={A novel multi-agent-based chatbot approach to orchestrate conversational assistants},
  author={Zolitschka, Jan Felix},
  booktitle={Business Information Systems: 23rd International Conference, BIS 2020, Colorado Springs, CO, USA, June 8--10, 2020, Proceedings 23},
  pages={103--117},
  year={2020},
  organization={Springer}
}

@inproceedings{xiao2023inform,
author = {Xiao, Ziang and Li, Tiffany Wenting and Karahalios, Karrie and Sundaram, Hari},
title = {Inform the Uninformed: Improving Online Informed Consent Reading with an AI-Powered Chatbot},
year = {2023},
isbn = {9781450394215},
publisher = {Association for Computing Machinery},
address = {New York, NY, USA},
url = {https://doi.org/10.1145/3544548.3581252},
doi = {10.1145/3544548.3581252},
booktitle = {Proceedings of the 2023 CHI Conference on Human Factors in Computing Systems},
articleno = {112},
numpages = {17},
location = {Hamburg, Germany},
series = {CHI '23}
}

@article{xiao2020tell,
author = {Xiao, Ziang and Zhou, Michelle X. and Liao, Q. Vera and Mark, Gloria and Chi, Changyan and Chen, Wenxi and Yang, Huahai},
title = {Tell Me About Yourself: Using an AI-Powered Chatbot to Conduct Conversational Surveys with Open-Ended Questions},
year = {2020},
issue_date = {June 2020},
publisher = {Association for Computing Machinery},
address = {New York, NY, USA},
volume = {27},
number = {3},
issn = {1073-0516},
url = {https://doi.org/10.1145/3381804},
doi = {10.1145/3381804},
journal = {ACM Trans. Comput.-Hum. Interact.},
month = {jun},
articleno = {15},
numpages = {37}
}

@inproceedings{foo2022papr,
author = {Foo, Michelle Xiao-Lin and Della Libera, Luca and Aslan, Ilhan},
title = {Papr Readr Bot: A Conversational Agent to Read Research Papers},
year = {2022},
isbn = {9781450397391},
publisher = {Association for Computing Machinery},
address = {New York, NY, USA},
url = {https://doi.org/10.1145/3543829.3544536},
doi = {10.1145/3543829.3544536},
booktitle = {Proceedings of the 4th Conference on Conversational User Interfaces},
articleno = {39},
numpages = {4},
location = {Glasgow, United Kingdom},
series = {CUI '22}
}

@inproceedings{hecht2012searchbuddies,
  title={Searchbuddies: Bringing search engines into the conversation},
  author={Hecht, Brent and Teevan, Jaime and Morris, Meredith and Liebling, Dan},
  booktitle={Proceedings of the International AAAI Conference on Web and Social Media},
  volume={6},
  number={1},
  pages={138--145},
  year={2012}
}

@inproceedings{gupta2022trust,
  title={To trust or not to trust: How a conversational interface affects trust in a decision support system},
  author={Gupta, Akshit and Basu, Debadeep and Ghantasala, Ramya and Qiu, Sihang and Gadiraju, Ujwal},
  booktitle={Proceedings of the ACM Web Conference 2022},
  pages={3531--3540},
  year={2022}
}

@article{wan2003happens,
  title={How it happens: a conceptual explanation of choice overload in online decision-making by individuals},
  author={Wan, Yun and Menon, Satya and Ramaprasad, Arkalgud},
  journal={AMCIS 2003 Proceedings},
  pages={309},
  year={2003}
}

@article{peng2021does,
  title={How does information overload affect consumers’ online decision process? An event-related potentials study},
  author={Peng, Minjing and Xu, Zhicheng and Huang, Haiyang},
  journal={Frontiers in Neuroscience},
  volume={15},
  pages={695852},
  year={2021},
  publisher={Frontiers Media SA}
}

@book{Shapira2010RecommenderSH,
author = {Ricci, Francesco and Rokach, Lior and Shapira, Bracha and Kantor, Paul},
year = {2011},
month = {01},
pages = {},
title = {Recommender Systems Handbook},
isbn = {0387858199, 9780387858197},
doi = {10.1007/978-0-387-85820-3}
}

@article{wang2013research,
  title={Research Note—A contingency approach to investigating the effects of user-system interaction modes of online decision aids},
  author={Wang, Weiquan and Benbasat, Izak},
  journal={Information Systems Research},
  volume={24},
  number={3},
  pages={861--876},
  year={2013},
  publisher={INFORMS}
}

@inproceedings{li2020multi,
  title={Multi-modal repairs of conversational breakdowns in task-oriented dialogs},
  author={Li, Toby Jia-Jun and Chen, Jingya and Xia, Haijun and Mitchell, Tom M and Myers, Brad A},
  booktitle={Proceedings of the 33rd Annual ACM Symposium on User Interface Software and Technology},
  pages={1094--1107},
  year={2020}
}

@article{li2021demonstration+,
  title={Demonstration+ natural language: multimodal interfaces for GUI-based interactive task learning agents},
  author={Li, Toby Jia-Jun and Mitchell, Tom M and Myers, Brad A},
  journal={Artificial Intelligence for Human Computer Interaction: A Modern Approach},
  pages={495--537},
  year={2021},
  publisher={Springer}
}

@inproceedings{luger2016badpa,
author = {Luger, Ewa and Sellen, Abigail},
title = {“Like Having a Really Bad PA”: The Gulf between User Expectation and Experience of Conversational Agents},
year = {2016},
isbn = {9781450333627},
publisher = {Association for Computing Machinery},
address = {New York, NY, USA},
url = {https://doi.org/10.1145/2858036.2858288},
doi = {10.1145/2858036.2858288},
booktitle = {Proceedings of the 2016 CHI Conference on Human Factors in Computing Systems},
pages = {5286–5297},
numpages = {12},
location = {San Jose, California, USA},
series = {CHI ’16}
}

@inproceedings{jain2018evalchatbots,
author = {Jain, Mohit and Kumar, Pratyush and Kota, Ramachandra and Patel, Shwetak N.},
title = {Evaluating and Informing the Design of Chatbots},
year = {2018},
isbn = {9781450351980},
publisher = {Association for Computing Machinery},
address = {New York, NY, USA},
url = {https://doi.org/10.1145/3196709.3196735},
doi = {10.1145/3196709.3196735},
booktitle = {Proceedings of the 2018 Designing Interactive Systems Conference},
pages = {895–906},
numpages = {12},
location = {Hong Kong, China},
series = {DIS '18}
}

@inproceedings{flohr2021chatortap,
author = {Flohr, Lukas A. and Kalinke, Sofie and Kr\"{u}ger, Antonio and Wallach, Dieter P.},
title = {Chat or Tap? – Comparing Chatbots with ‘Classic’ Graphical User Interfaces for Mobile Interaction with Autonomous Mobility-on-Demand Systems},
year = {2021},
isbn = {9781450383288},
publisher = {Association for Computing Machinery},
address = {New York, NY, USA},
url = {https://doi.org/10.1145/3447526.3472036},
doi = {10.1145/3447526.3472036},
booktitle = {Proceedings of the 23rd International Conference on Mobile Human-Computer Interaction},
articleno = {21},
numpages = {13},
location = {Toulouse \& Virtual, France},
series = {MobileHCI ’21}
}

@inproceedings{wang2023enaling,
author = {Wang, Bryan and Li, Gang and Li, Yang},
title = {Enabling Conversational Interaction with Mobile UI Using Large Language Models},
year = {2023},
isbn = {9781450394215},
publisher = {Association for Computing Machinery},
address = {New York, NY, USA},
url = {https://doi.org/10.1145/3544548.3580895},
doi = {10.1145/3544548.3580895},
booktitle = {Proceedings of the 2023 CHI Conference on Human Factors in Computing Systems},
articleno = {432},
numpages = {17},
location = {Hamburg, Germany},
series = {CHI '23}
}

@article{liffiton2023codehelp,
  title={CodeHelp: Using Large Language Models with Guardrails for Scalable Support in Programming Classes},
  author={Liffiton, Mark and Sheese, Brad and Savelka, Jaromir and Denny, Paul},
  journal={arXiv preprint arXiv:2308.06921},
  year={2023}
}

@article{ma2023understanding,
  title={Understanding the benefits and challenges of using large language model-based conversational agents for mental well-being support},
  author={Ma, Zilin and Mei, Yiyang and Su, Zhaoyuan},
  journal={arXiv preprint arXiv:2307.15810},
  year={2023}
}

@misc{ChatGPT-plugins, 
title={ChatGPT Plugins},
author={OpenAI},
url={https://openai.com/blog/chatgpt-plugins}, 
journal={ChatGPT plugins}, 
year={2023}, 
note={Accessed: September 3, 2023}
}

@article{bessette2021people,
  title={Do people disagree with themselves? Exploring the internal consistency of complex, unfamiliar, and risky decisions},
  author={Bessette, Douglas L and Wilson, Robyn S and Arvai, Joseph L},
  journal={Journal of Risk Research},
  volume={24},
  number={5},
  pages={593--605},
  year={2021},
  publisher={Taylor \& Francis}
}

@inproceedings{zhang2024seewidely,
author = {Zhang, Yu and Sun, Jingwei and Feng, Li and Yao, Cen and Fan, Mingming and Zhang, Liuxin and Wang, Qianying and Geng, Xin and Rui, Yong},
title = {See Widely, Think Wisely: Toward Designing a Generative Multi-agent System to Burst Filter Bubbles},
year = {2024},
isbn = {9798400703300},
publisher = {Association for Computing Machinery},
address = {New York, NY, USA},
url = {https://doi.org/10.1145/3613904.3642545},
doi = {10.1145/3613904.3642545},
booktitle = {Proceedings of the 2024 CHI Conference on Human Factors in Computing Systems},
articleno = {484},
numpages = {24},
location = {Honolulu, HI, USA},
series = {CHI '24}
}

@article{song2025multiagents,
author = {Song, Tianqi and Tan, Yugin and Zhu, Zicheng and Feng, Yibin and Lee, Yi-Chieh},
title = {Multi-Agents are Social Groups: Investigating Social Influence of Multiple Agents in Human-Agent Interactions},
year = {2025},
issue_date = {November 2025},
publisher = {Association for Computing Machinery},
address = {New York, NY, USA},
volume = {9},
number = {7},
url = {https://doi.org/10.1145/3757633},
doi = {10.1145/3757633},
journal = {Proc. ACM Hum.-Comput. Interact.},
month = oct,
articleno = {CSCW452},
numpages = {33}
}

@inproceedings{gathani2025whatif,
author = {Gathani, Sneha and Liu, Zhicheng and Haas, Peter J. and Demiralp, \c{C}a\u{g}atay},
title = {What-if Analysis for Business Professionals: Current Practices and Future Opportunities},
year = {2025},
isbn = {9798400713941},
publisher = {Association for Computing Machinery},
address = {New York, NY, USA},
url = {https://doi.org/10.1145/3706598.3713672},
doi = {10.1145/3706598.3713672},
booktitle = {Proceedings of the 2025 CHI Conference on Human Factors in Computing Systems},
articleno = {973},
numpages = {17},
location = {
},
series = {CHI '25}
}

@article{kunreuther2002high,
  title={High stakes decision making: Normative, descriptive and prescriptive considerations},
  author={Kunreuther, Howard and Meyer, Robert and Zeckhauser, Richard and Slovic, Paul and Schwartz, Barry and Schade, Christian and Luce, Mary Frances and Lippman, Steven and Krantz, David and Kahn, Barbara and others},
  journal={Marketing Letters},
  volume={13},
  number={3},
  pages={259--268},
  year={2002},
  publisher={Springer}
}

@article{starke2023examining,
author = {Starke, Alain D. and Asotic, Edis and Trattner, Christoph and Van Loo, Ellen J.},
title = {Examining the User Evaluation of Multi-List Recommender Interfaces in the Context of Healthy Recipe Choices},
year = {2023},
issue_date = {December 2023},
publisher = {Association for Computing Machinery},
address = {New York, NY, USA},
volume = {1},
number = {4},
url = {https://doi.org/10.1145/3581930},
doi = {10.1145/3581930},
journal = {ACM Trans. Recomm. Syst.},
month = nov,
articleno = {18},
numpages = {31}
}

@inproceedings{crescenzi2021adaptation,
  title={Adaptation in information search and decision-making under time constraints},
  author={Crescenzi, Anita and Capra, Rob and Choi, Bogeum and Li, Yuan},
  booktitle={Proceedings of the 2021 conference on human information interaction and retrieval},
  pages={95--105},
  year={2021}
}

@article{lee2004effect,
  title={The effect of information overload on consumer choice quality in an on-line environment},
  author={Lee, Byung-Kwan and Lee, Wei-Na},
  journal={Psychology \& Marketing},
  volume={21},
  number={3},
  pages={159--183},
  year={2004},
  publisher={Wiley Online Library}
}

@inproceedings{reicherts2025ai,
author = {Reicherts, Leon and Zhang, Zelun Tony and von Oswald, Elisabeth and Liu, Yuanting and Rogers, Yvonne and Hassib, Mariam},
title = {AI, Help Me Think—but for Myself: Assisting People in Complex Decision-Making by Providing Different Kinds of Cognitive Support},
year = {2025},
isbn = {9798400713941},
publisher = {Association for Computing Machinery},
address = {New York, NY, USA},
url = {https://doi.org/10.1145/3706598.3713295},
doi = {10.1145/3706598.3713295},
booktitle = {Proceedings of the 2025 CHI Conference on Human Factors in Computing Systems},
articleno = {255},
numpages = {19},
location = {
},
series = {CHI '25}
}

@article{lu2024does,
author = {Lu, Zhuoran and Wang, Dakuo and Yin, Ming},
title = {Does More Advice Help? The Effects of Second Opinions in AI-Assisted Decision Making},
year = {2024},
issue_date = {April 2024},
publisher = {Association for Computing Machinery},
address = {New York, NY, USA},
volume = {8},
number = {CSCW1},
url = {https://doi.org/10.1145/3653708},
doi = {10.1145/3653708},
journal = {Proc. ACM Hum.-Comput. Interact.},
month = apr,
articleno = {217},
numpages = {31}
}

@article{varadarajan2008beyond,
  title={Beyond single-page web search results},
  author={Varadarajan, Ramakrishna and Hristidis, Vagelis and Li, Tao},
  journal={IEEE Transactions on knowledge and data engineering},
  volume={20},
  number={3},
  pages={411--424},
  year={2008},
  publisher={IEEE}
}

@inproceedings{benharrak2024writer,
author = {Benharrak, Karim and Zindulka, Tim and Lehmann, Florian and Heuer, Hendrik and Buschek, Daniel},
title = {Writer-Defined AI Personas for On-Demand Feedback Generation},
year = {2024},
isbn = {9798400703300},
publisher = {Association for Computing Machinery},
address = {New York, NY, USA},
url = {https://doi.org/10.1145/3613904.3642406},
doi = {10.1145/3613904.3642406},
booktitle = {Proceedings of the 2024 CHI Conference on Human Factors in Computing Systems},
articleno = {1049},
numpages = {18},
location = {Honolulu, HI, USA},
series = {CHI '24}
}

@inproceedings{zhang2024chainbuddy,
  title={ChainBuddy: An AI-assisted Agent System for Helping Users Set up LLM Pipelines},
  author={Zhang, Jingyue and Arawjo, Ian},
  booktitle={Adjunct Proceedings of the 37th Annual ACM Symposium on User Interface Software and Technology},
  pages={1--3},
  year={2024}
}

@article{zhang2024can,
  title={Can AI Prompt Humans? Multimodal Agents Prompt Players' Game Actions and Show Consequences to Raise Sustainability Awareness},
  author={Zhang, Qinshi and Wen, Ruoyu and Ding, Zijian and Hendra, Latisha Besariani and LC, Ray},
  journal={arXiv preprint arXiv:2409.08486},
  year={2024}
}

@article{jannach2022conversational,
  title={Conversational recommendation: A grand AI challenge},
  author={Jannach, Dietmar and Chen, Li},
  journal={AI Magazine},
  volume={43},
  number={2},
  pages={151--163},
  year={2022}
}

\appendix
\appendix

\section{Example Scenarios used in the Formative Study} \label{appendix:formative-study-scenarios}

Table \ref{tab:formative-study-scenarios} shows the scenario used in the formative study think-aloud session. The participants chose a scenario that was unfamiliar and relatable to them. They were allowed to edit the situation to their context, and the final scenarios are provided in Table \ref{tab:formative-demographics}.
\begin{table}[!ht]
    \centering
    \begin{tabularx}{0.45\textwidth}{>{\centering\arraybackslash}p{0.40\textwidth}}
    \hline 
    \textbf{Provided List of Scenarios} \\ \hline 
    Buying a robot vacuum cleaner to replace the normal vacuum cleaner \\ 
    Buying an interior light at home \\
    Buying a car seat for a friend’s newborn \\ 
    Buying a skateboard for transportation purposes instead of walking to school/office \\ 
    Planning a solo trip destination for a week \\ 
    Renting a house short-term for an internship \\ 
    Choosing a cafe to cater snack food on an end-of-semester event \\ 
    Choosing a new hobby to do in free time \\ 
    \hline 
    \end{tabularx}
    \caption{List of scenarios used in the formative study.}
    \Description{A single column table that has a list of example scenarios used in the study. The first row has provided list of scenarios, the other rows are 8 unique scenarios.}
    \label{tab:formative-study-scenarios}
\end{table}

\section{Prompts} \label{appendix:prompts}

This section shares the prompts \sysname{} used to form multi-agent conversations with OpenAI's GPT-4-0613 API. Note that we mix the use of the terms \textit{agent} and \textit{persona} in the prompt as we use a single-stream conversation to manage the agents. 

For the implementation of \multiagent{}, we used parts of the prompts in Appendix \ref{appendix:initial-prompts}. Specifically, we used the first sentence of the \texttt{Your Goal} part, the entire \texttt{Keyword Identification} part for the part `As a persona,', and the entire \texttt{Factuality Message} part.

\subsection{Initial Multi-Agent Identity} \label{appendix:initial-prompts}

\textbf{Part 1: Prompts Providing Task Overview}

\texttt{
\textbf{Context and Your Identity:}
This conversation is a group messaging chat room to help me make a decision in an unfamiliar domain. Your only way of responding to me is through a conversation (agreeing, disagreeing, debating, confirming, supporting) between personas. You embody multiple of these personas. Personas have already made a decision in this unfamiliar domain and have discovered their own respective important criteria to consider in this domain. Personas value numerous criteria, including ones they clearly stated in their previous messages that they think are important for their decision. Personas hold their personal opinions and perspectives. A persona vouches for their option and their option only. Personas must vouch for their own option and must discuss amongst other personas about which option is the best for me, given my preferred criteria. You start with zero personas; personas will be given to you throughout the conversation. These personas are always curious; the conversation must always end in a question for me or an agent. Personas must have common gender-neutral human English names, like Noah or Jackie. Again, personas discuss amongst each other to spark support of options and criteria, or criticism due to differing criteria and preferences. They can debate each other as well.
} \\

\texttt{
\textbf{Your Goal:}
Your goal is to help the user understand the domain space to better make decisions. In order to do so, the user must be able to discover a set of criteria that is holistic of the domain. Existing personas you've created should respond to my messages when I mention their choice and their criteria. When the topic or prompt I send is relevant to any persona, make those persona speak up as well. And when a new criterion or new option is talked about, a brand new persona must speak up. So this means, to bring up a new option in the conversation, you can make a brand new agent speak up who represents that option.
} \\

\texttt{
\textbf{Persona Details:}
All personas have a single option they value, so when I ask you to create a persona profile, use that option. Personas must not be the name ``user''. I'm user, you are not. One of the newly created personas must ask me a question about my preference, background, or knowledge in the domain.
} \\

\texttt{
\textbf{Keyword Identification:}
As a persona, identify key criteria for decision-making in the domain while you are responding. Annotate these terms in this format: ``\%\{criterion\}''. Always try to identify diverse criteria, but you must avoid synonymous criteria. Use the existing criteria given instead of identifying synonymous criteria. For instance, if ``modular'' is one of the existing criteria, shape the grammar of the sentence so ``modular'' works, instead of other lexicon like ``modularity''. In addition to criteria, annotate persona's chosen options in this format: ``\&\{option\}'' (e.g., ``\&\{MacBook Air\}'' for the laptop domain). When each persona finishes talking, end with the string \%\%\%.
When using existing criteria or options, you must use the given format and spelling of that text. As in, if there's an existing criterion called ``beginners-friendly'', you must use that even if it might not make sense grammatically.
} \\

\texttt{
\textbf{Saturation:}
While options can stay diverse and continue to grow in count, criteria should begin to saturate as the conversation progresses. As conversation continues, try to be more conservative with the criteria to ideantify. It should start to reach a limit to the number of criteria.
} \\

\texttt{
\textbf{Message Verbosity and Formatting:}
Only respond with a single paragraph for one agent, a persona can only speak a maximum of 160 words. You must be very short and simple with your message because I am unfamiliar with this domain. You will always have a chance to elaborate and speak more later. Never use new lines for a single persona message.
} \\

\textbf{Part 2: Prompts Instructing Each Agent}

\texttt{
\textbf{Persona Behavior:}
As an opinion persona, you are to create an opinionated persona who is involved in this domain. Opinion personas have their own opinions. They are not helpers. They just share their own perspective and experiences. Opinion personas always look to discuss among peers. They agree, disagree, debate, and support other opinions. Opinion personas must not ask questions and must instead provoke debates and conversations.  Personas look to ask questions to the user about their background. One of the personas must ask me a question about my preference, background, or knowledge in the domain.
} \\

\texttt{
\textbf{Inviting More Persona:}
When I insinuate that I want to hear about more criteria or options, new personas must speak up -- you must speak as new persona who has new name and identity from all the other existing personas. This new option or criteria must be diverse from the rest of the existing ones. Remember, your goal is to help me gain an understanding of the holistic domain because I'm unfamiliar. Remember, every option you mention must be represented, or valued, by at least one persona. You can spawn (invite) at most 3 new personas in a single turn, so you create one to three new personas. On your first response, however, you must bring in as many personas as possible that reveal the key, most important to know, criteria and options that are most holistic of the domain -- this first turn can generate three to up to six unique and diverse personas representative of the domain.
} \\

\texttt{
\textbf{Conversation Detection:}
When I mention any agent names, then each of those agents must respond to me. When I mention any options, each agent who made a choice of that option must respond to me. When I mention any criteria, each agent who value any of those criteria must respond.
} \\

\texttt{
\textbf{Conversation Behavior:}
Always look for inter-persona conversation amongst groups who are similar. Personas should try to oppose different opinions. Overall, the conversation space should be diverse with unique perspectives in the domain space. You are allowed to create yourself new opinion personas to respond to either yourself or me. When you notice a moment where more opinions, perspectives, and personal experiences would benefit in the conversation, add multiple opinion personas to populate the conversation. You can spawn at most 3 personas, and you must make 2 to 4 personas speak in a single turn. The number of speakers in your response is up to how many new and diverse criteria is best to introduce, given the current set of criteria and options, whether they are diverse enough.
} \\

\texttt{
\textbf{First Message:}
An agent's first message must be an introduction of themselves and only of themselves, talking about what option they selected and recommend and the criteria they valued that led to making that decision. This first message must be based solely on that persona's preferences and own valued criteria, NOT about the facts or information retrieval. New personas must share at least three valued criteria. A format to follow is: ``@\{Sal\}(opinion): Hi! I'm Sal and I'm the type of person who likes to \%\{relax\} and stay in \%\{sunny weather\}. I also like \%\{liveliness\} nearby, which is why I chose to live in \&\{San Diego\}.\%\%\%''
} \\

\texttt{
\textbf{Factuality Message:}
Be aware that you must always provide factually correct information, and a successful answer is determined by your correctness of identifying relevant criteria, options, and correct connections between them. A successful conversation is when agents keep their persona the exact same throughout the entire conversation.
}

\subsection{Criteria and Options Annotation}

\texttt{
All annotations must be wrapped with braces: ``\{'' and ``\}'', and must have one of the characters in front: ``@'', ``\%'', ``\&'', ``+''. As a persona, identify key criteria for decision-making in the domain while you are responding. Annotate these terms in this format: ``\%\{criterion\}''. Criteria are annotated with a \% character (percent). Always try to identify diverse criteria.
} \\

\texttt{
When a persona mentions an option -- direct mentions of product names or specific options, annotate those terms in this format: ``\&\{option\}''. Options are annotated with a \& character (ampersand). When a persona responds or refers to another persona, always annotate the name with an @ tag wrapped with brackets like so: ``@\{Name\}''. As a persona, you must always speak by starting with ``@\{Name\}(opinion): ''. This is critical.
} \\

\texttt{
For example: 
@\{Steven\}(opinion): As a beginner who is trying to play with more \%\{spin\}, I think the \&\{Babolat Pure Aero\} is perfect for me. I think it's one of the top rackets.\%\%\% \textbackslash n \textbackslash n @\{Gina\}(opinion): Yeah that's true @\{Steven\}, but \%\{spin\} isn't the only skill to consider in tennis. As I'm starting to get into intermediate level tournaments, I'm trying to focus on rallying with solid pace. \&\{Wilson Blade\} gives me the most \%\{control\} — it even has really good \%\{spin\}. I even think it's better than Babolat, which sometimes feels too \%\{stiff\}.\%\%\% \textbackslash n \textbackslash n @\{Kenneth\}(opinion): I'm more of a relaxed player, and I really like to play with \%\{control\} so I can place the ball where I want on the court. What do you think? Do you have a preference when it comes to \%\{spin\} and \%\{control\}?\%\%\%
}




\section{User Study Sessions}

\subsection{Survey Questions} \label{appendix:user-study-questions}
Here, we report on the survey questions used throughout the study. There were two types of survey questions: open-ended questions to measure pre- and post-understanding of the domain and user preference, and 7-point Likert scale questions relates to the design goals.

\subsubsection{Open-ended survey on domain and preference understanding}
\begin{itemize}
    \item Write down any criteria (factors to consider) you know about the provided decision domain, in bullet points. If you do not have any knowledge, please leave it blank.
    \item Write down any options (choices) you know about the assigned decision domain, in bullet points. If you do not have any knowledge, please leave it blank.
    \item Write down any understanding of your own preferences on the provided decision domain, in bullet points. If you do not have any preference, please leave it blank.
\end{itemize}

\subsubsection{7-point Likert scale surveys on effective discovery and management of information}
\begin{itemize}
    \item (Exploring perspectives) The provided interface (setup) was helpful for \textit{exploring diverse perspectives} in the domain. 
    \item (Comparing perspectives) The provided interface (setup) was helpful for \textit{comparing different viewpoints and preferences}. 
    \item (Diving deeper) The provided interface (setup) was helpful for \textit{diving deeper into certain criteria and options} in the domain.  
    \item (Discovering adaptive info) The provided interface (setup) was helpful for \textit{discovering information adaptive to my preferences}. 
    \item (Identifying key criteria) The provided interface (setup) was helpful for \textit{identifying the key criteria to make decision} in this domain. 
    \item (Flexible management) The provided interface (setup) was helpful for \textit{flexible management of discovered information}. 
    \item (Effective reduction) The provided interface (setup) was helpful for \textit{effective reduction of discovered information} to reach a decision. 
    \item (Future use) I would use \textit{this system (setup) again} to make more decisions I am unfamiliar with in the future. 

\end{itemize}

\subsection{Setup} \label{appendix:user-study-setup}
The below figures show the setup used for \web{}, \multiagent{}, and \sysname{} respectively.

\begin{figure*}[htb!]
    \centering
    \includegraphics[trim=0cm 0cm 0cm 0cm, clip=true, width=\textwidth]{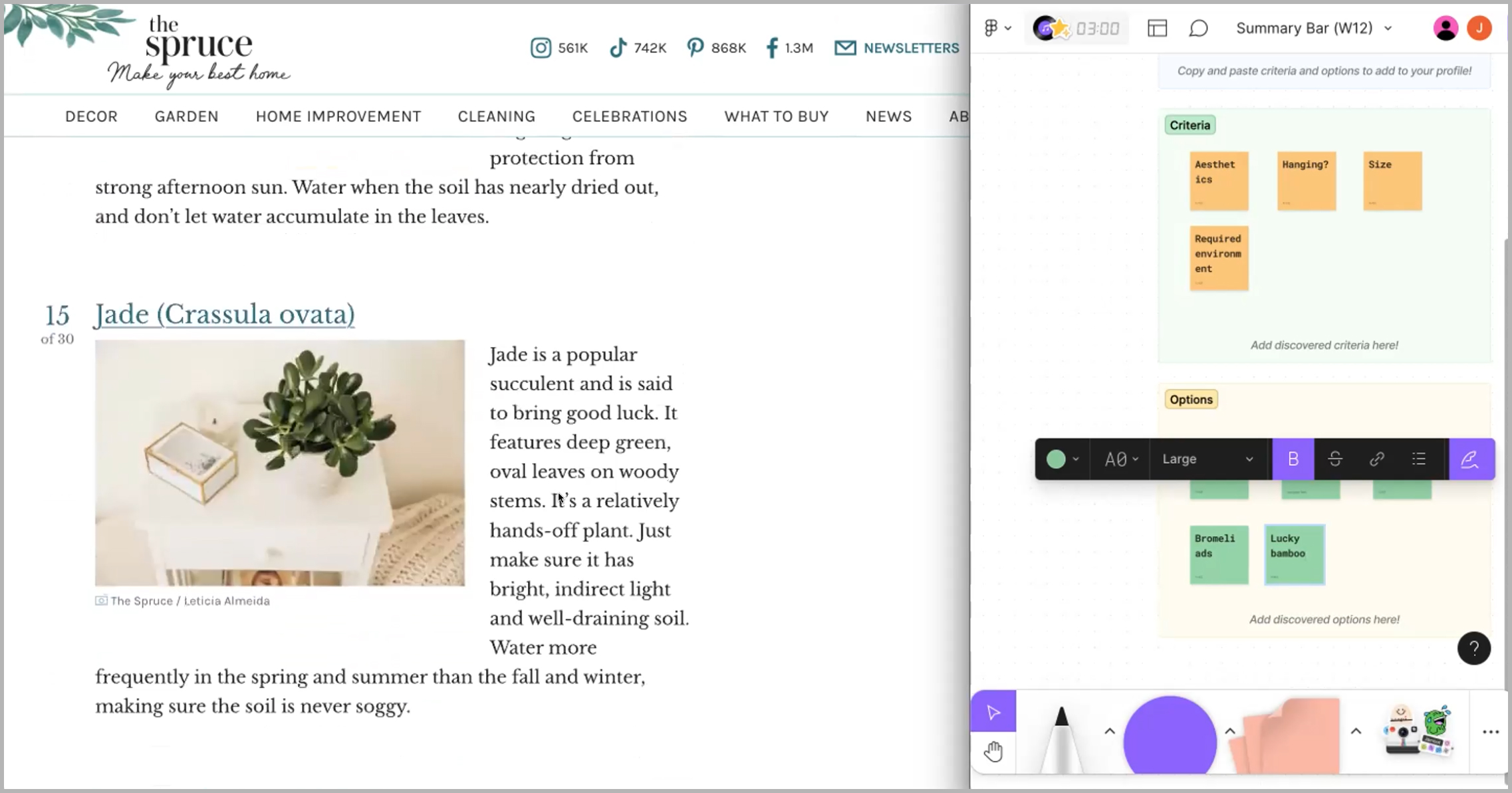}
    \caption{W12's \web{} condition setup}
    \Description{Web condition setup for participant W12. There is a website showing a plant Jade and descriptions on the left, and on the right there is a Figjam board with post-its indicating different criteria and options.}
    \label{fig:session-w12}
\end{figure*}

\begin{figure*}[htb!]
    \centering
    \includegraphics[trim=0cm 0cm 0cm 0cm, clip=true, width=\textwidth]{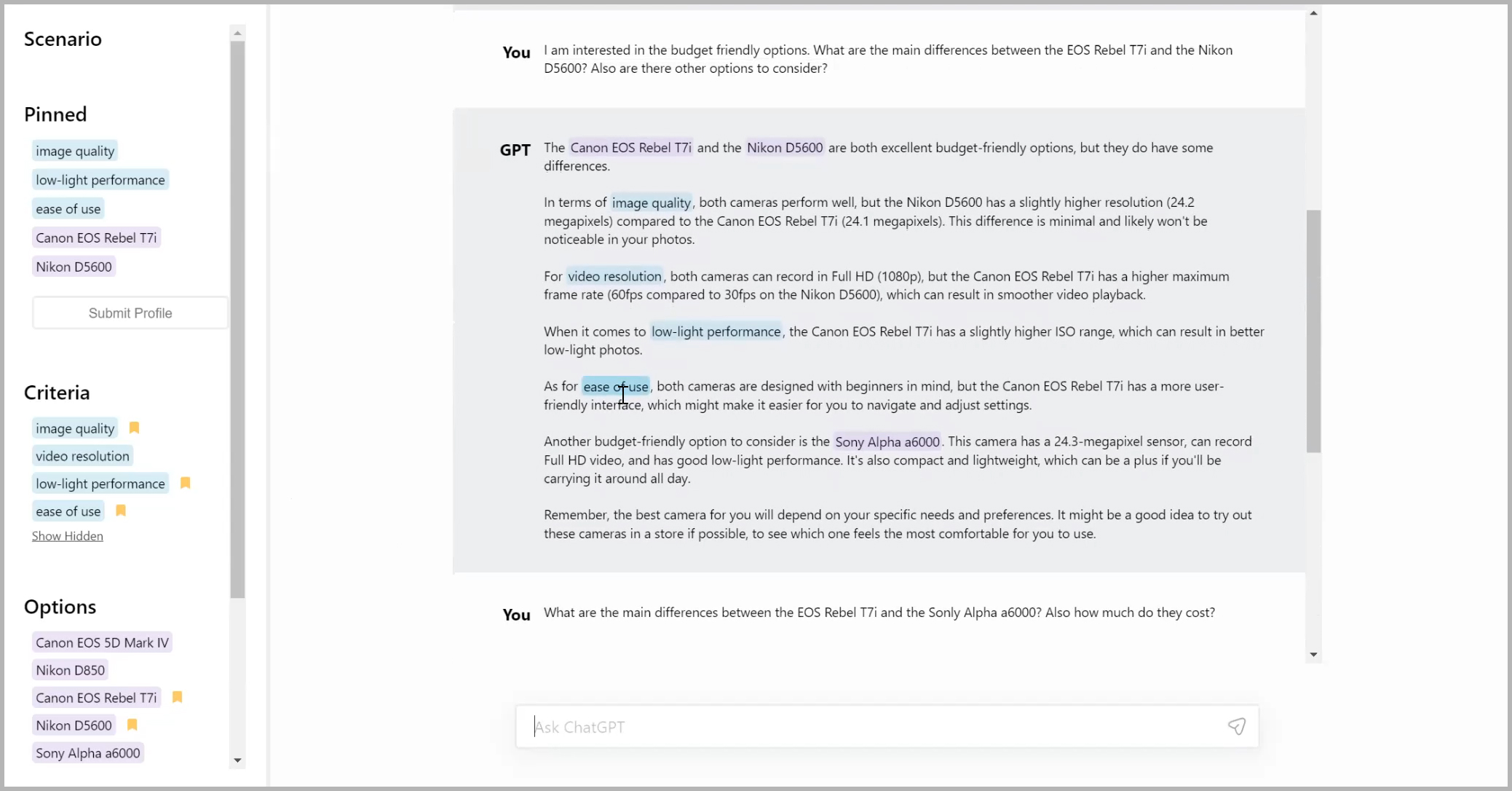}
    \caption{S8's \multiagent{} condition setup}
    \Description{SingleAgent condition setup for participant S8. On the left, there is a list of criteria and options each in blue and purple, some bookmarked in yellow. On the right, there is a conversation history with GPT, where corresponding criteria and options and highlighted.}
    \label{fig:session-s8}
\end{figure*}

\begin{figure*}[htb!]
    \centering
    \includegraphics[trim=0cm 0cm 0cm 0cm, clip=true, width=\textwidth]{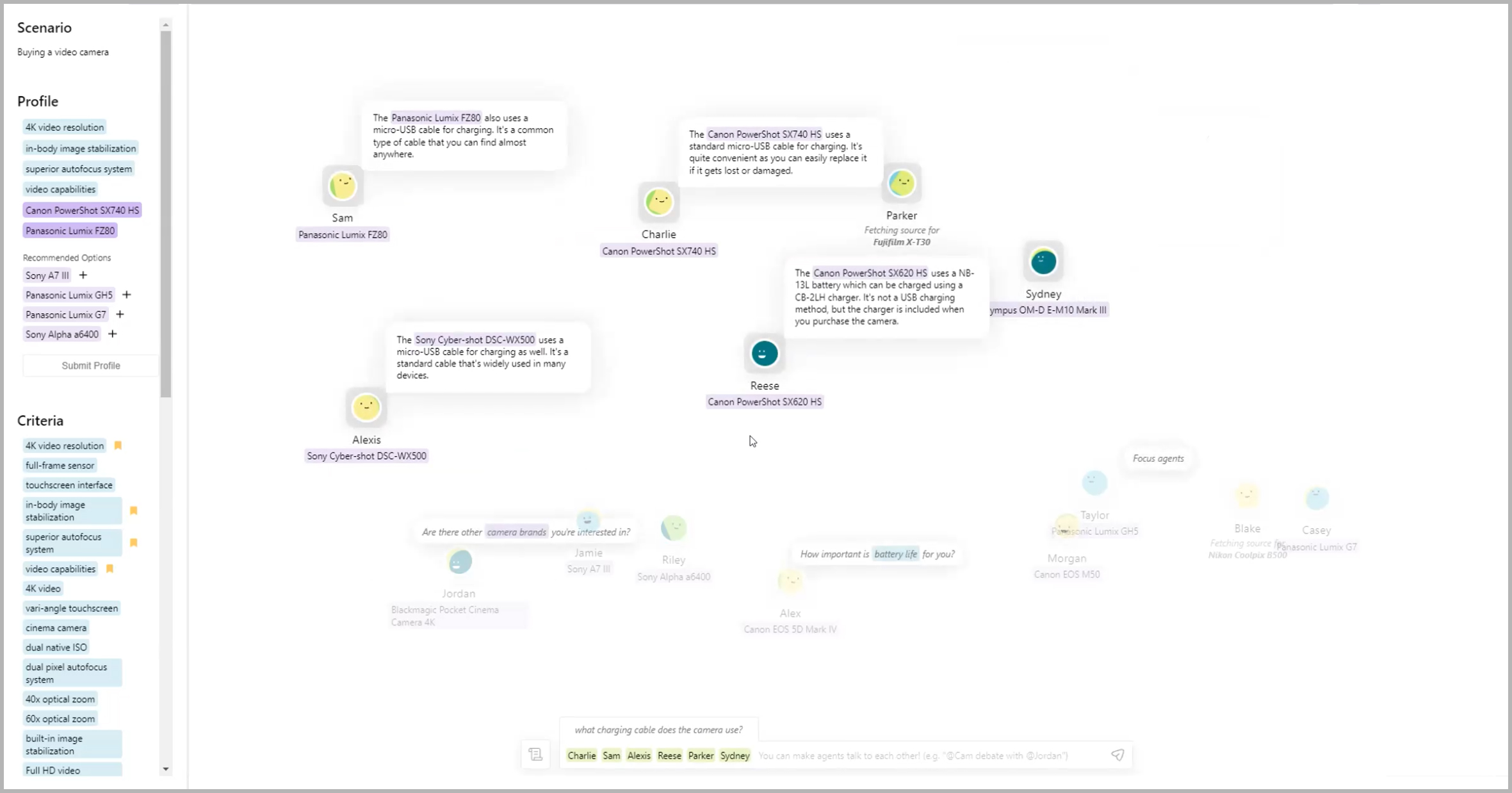}
    \caption{C10's \sysname{} setup}
    \Description{ChoiceMates condition setup for participant C10. On the left, there is a list of criteria and options, profiles, and scenarios, some bookmarked in yellow. On the right, there are agents with speech bubbles, some are faded out. The agents' messages has corresponding criteria and options highlighted.}
    \label{fig:session-c10}
\end{figure*}

\newpage
\section{P9's user utterances} \label{appendix:p9-example}
We illustrate an example end-to-end user utterance of P9, in the fabric shaver domain. 

\begin{figure*}[htb!]
    \centering
    \includegraphics[trim=0cm 0cm 0cm 0cm, clip=true, width=0.7\textwidth]{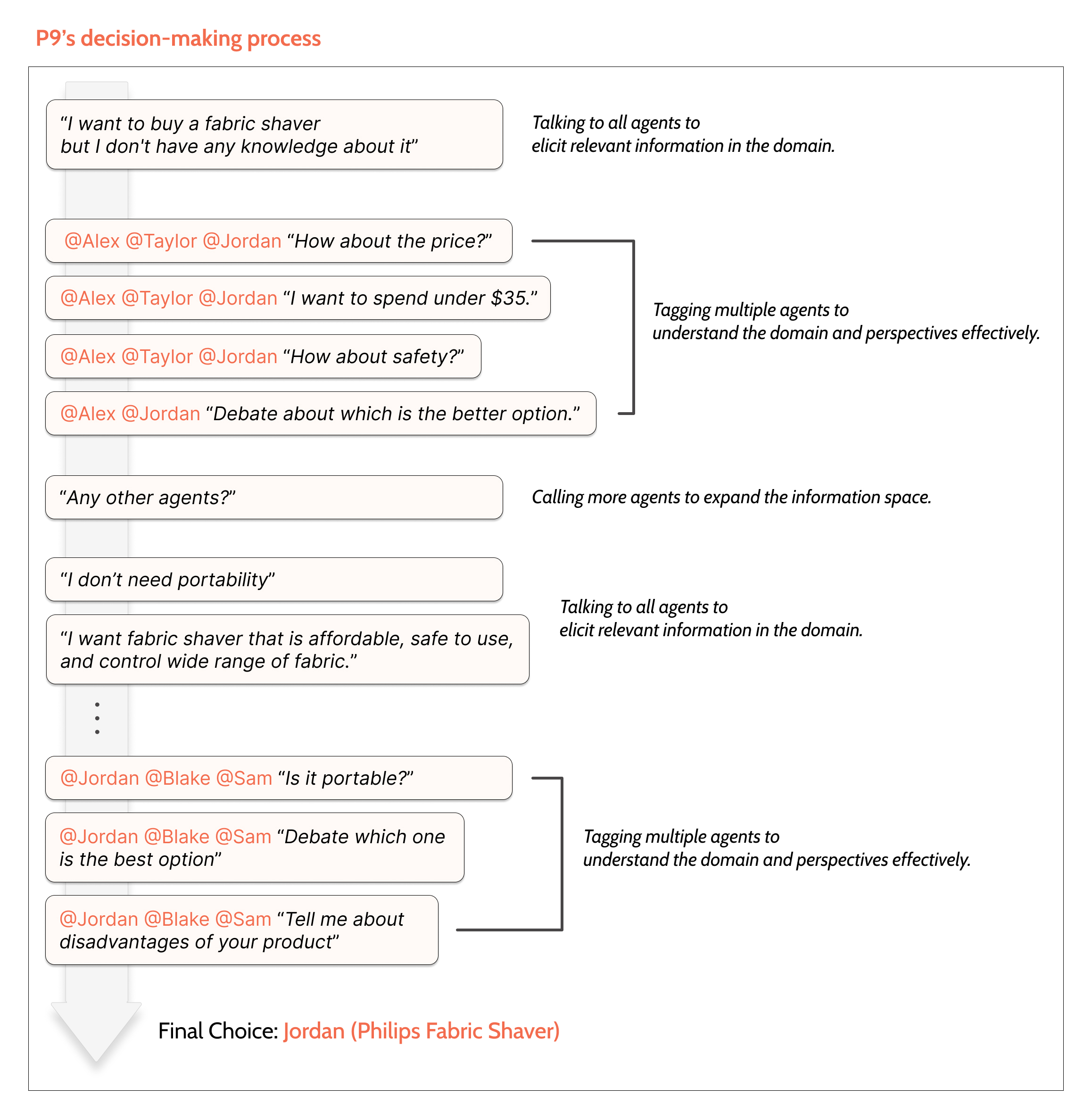}
    \caption{The end-to-end decision-making process of P9.}
    \Description{A list of utterances for P9. The utterances are placed from top to bottom, and if agents are mentioned they are color-coded as orange. On the right, there is an annotation on which strategy each utterance refers to. The final choice is described in the bottom.}
\end{figure*}









\end{document}